\documentclass[preprint,letterpaper,showpacs]{revtex4}
\usepackage{graphicx}
\def \zc {z_c}
\def \iv {\int d\nu \,}

\def \cG {{\cal G}}
\def \cuG {{\underline{\cal G}}}
\def \uGr {{\underline{G}}}
\def \Det {{\rm Det}}
\def \Tan {{\rm Tan}}
\def \re {{\rm Re}\,}
\def \im {{\rm Im}\,}

\def \vS {{\bf S}}

\def \vR {{\bf R}}

\def \uc {\underline{\chi }}
\def \uGm {\underline{\Gamma }}
\def \vk {{\bf k}}
\def \vq {{\bf q}}
\def \vQ {{\bf Q}}
\def \om {\omega_m}
\def \cel {\chi^{\rm el}}
\def \ds {\displaystyle}
\def \pt {\delta }
\def \al {\eta }
\def \bal {\bar{\eta }}

\def \hm {{\bf m}}
\def \Tr {{\rm Tr_{f}}}
\def \TC {T_{\rm C}}
\def \TN {T_{\rm N}}
\def \TM {T_{\rm SRO}}
\def \TR {T^{(\rm max )}_{\rm SRO}}
\def \Etmin {E^{({\rm min })}}

\def \bz {{\bf z}}

\def \bs {{\bf \sigma }}
\def \vS {{\bf S}}

\def \seff {S_{\rm eff}}
\def \uI {\underline{I}}
\def \uK {\underline{K}}
\def \vs {{\bf s}}
\def \sgn {{\rm sgn}}
\def \bc {\bar c}
\def \JH {J_{\rm H}}
\def \tJH {J}

\def \an {{\underline\Gamma}^{(0)}}

\begin{document}
\title{Magnetic Instabilities and Phase Diagram of the Double-Exchange Model in Infinite Dimensions}
\author{R S Fishman$^*$, F Popescu$^{\circ }$, G Alvarez$^{\bullet}$,
J Moreno$^{\dagger }$, Th Maier$^{\bullet}$, and M Jarrell$^{\ddagger }$}
\affiliation{$^*$Materials Science and Technology Division, Oak Ridge National Laboratory, Oak Ridge, TN 37831-6032}
\affiliation{$^{\circ }$Physics Department, Florida State University, Tallahassee,FL, 32306}
\affiliation{$^{\bullet}$Computer Science and Mathematics Division, Oak Ridge National Laboratory, Oak Ridge, TN 37831-6032}
\affiliation{$^{\dagger }$Physics Department, University of North Dakota, Grand Forks, ND 58202-7129}
\affiliation{$^{\ddagger }$Deparment of Physics, University of Cincinnati, Cincinnati, OH 45221}

\begin{abstract}

Dynamical mean-field theory is used to study the magnetic instabilities and phase
diagram of the double-exchange model with Hund's coupling $\JH >0$ in infinite dimensions.
In addition to ferromagnetic (FM) and antiferromagnetic (AF) phases, the DE model also supports a broad
class of short-range ordered (SRO) states with extensive entropy and short-range magnetic order.
For any site on the Bethe lattice, the correlation parameter $q$ of a SRO state is given by the average
$q=\langle \sin^2 (\theta_i/2)\rangle $, where $\theta_i$ is the angle between any spin and its
neighbors.  Unlike the FM ($q=0$) and AF ($q=1$) transitions, the transition temperature of a
SRO state with $0< q < 1$ cannot be obtained from the magnetic susceptibility.  But a solution of
the coupled Green's functions in the weak-coupling limit indicates that a SRO state always has a
higher transition temperature than the AF for all fillings $p$ below 1 and even has a higher
transition temperature than the FM for $0.26 \le p \le 0.39$.  For $0.39 < p < 0.73$, where
both the FM and AF phases are unstable for small $\JH $, a SRO phase has a non-zero transition
temperature except close to $p=0.5$.   As $\JH $ increases, the SRO transition
temperature eventually vanishes and the FM phase dominates the phase diagram.  For
small $\JH $, the $T=0$ phase diagram of the DE model is greatly simplified by
the presence of the SRO phase.  A SRO phase is found to have lower energy than
either the FM or AF phases for $0.26 \le p < 1$.  Phase separation disappears as $\JH \rightarrow 0$
but appears for any non-zero coupling.  For fillings near $p=1$, phase separation occurs between an AF with
$p=1$ and either a SRO or a FM phase.  The stability of a SRO state at $T=0$ can be understood by examining
the interacting density-of-states, which is gapped for any nonzero $\JH $ in an AF but only when
$\JH $ exceeds a critical value in a SRO state.

\end{abstract}

\maketitle

\section{Introduction}

With the renewed interest in manganites \cite{mang} and the revitalized study of dilute magnetic semiconductors
\cite{dms}, attention has once again focused on the double-exchange (DE) model and its variants.
Applied to the manganites, the DE model is customarily formulated with one local moment per site and a Hund's
coupling $\JH $ taken to be much larger than the electron bandwidth $W$.  In studies of dilute magnetic
semiconductors, on the other hand, the local moments are sparse and the exchange coupling
is comparable to $W$.  While the full phase diagram of the DE model as a function of $\JH $ and
electron filling $p$ has been studied by several authors \cite{yun:98,nag:00,cha:00,yin:03},
the properties of the DE model for small coupling constant remain very much in doubt.
In particular, there have been conflicting claims about the presence of phase separation (PS) \cite{cha:00}
and incommensurate phases \cite{yun:98,nag:00,yin:03} for small $\JH $, when the ordered
FM and AF phases are magnetically frustrated by a RKKY-like interaction between the local moments
\cite{van:62}.

This paper uses DMFT to evaluate the magnetic instabilities and $T=0$ phase diagram of the
DE model.   Developed in the late 1980's by M\"uller-Hartmann \cite{mul:89} and Metzner and Vollhardt \cite{met:89},
DMFT exploits the fact that the self-energy becomes independent of momentum in infinite dimensions,
where DMFT becomes formally exact.  Even in three dimensions, DMFT is believed to capture the physics of
correlated systems including the narrowing of electron bands and the Mott-Hubbard transition \cite{geo:96}.
Although DMFT has been widely applied to the DE model \cite{nag:00,cha:00,fur:95,mil:96,fur:99,aus:01,
fis:03,che:03,fis:05,ary:05}, until now there has been no complete treatment of the phase instabilities and
$T=0$ phase diagram of the DE model for arbitrary $\JH $ and $p$.

\begin{figure}
\includegraphics *[scale=0.8]{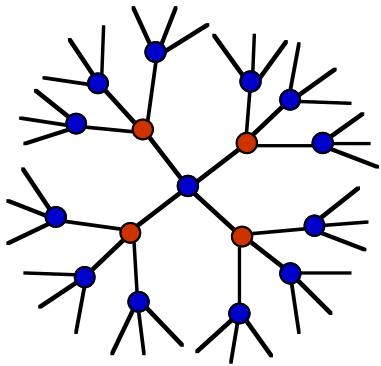}
\caption{
A Bethe lattice with $\zc =4$ nearest neighbors.  As shown, the Bethe lattice may be
partitioned into $A$ and $B$ sublattices, denoted by the blue and red dots.
}
\end{figure}

We shall study a system with a bare semicircular density-of-states (DOS) given by
$N_0(\omega )=(8/\pi W^2)\re \sqrt{W^2/4-\omega^2}$.  In real space, this DOS belongs
to an infinite-dimensional  Bethe lattice or an infinite Cayley tree with no closed loops \cite{bri:70}.
A finite-dimensional Bethe lattice with coordination number $\zc  =4$ is sketched
in Fig.1.  Although the Bethe lattice lacks translational symmetry. it is quite convenient for calculations.
As shown in Fig.1, the Bethe lattice can be partitioned into $A$ and $B$
sublattices so that both ferromagnetic (FM) and antiferromagnetic (AF) long-range order are possible.
Due to the bounds $\pm W/2$, the DOS of the Bethe lattice more closely resembles the DOS of two-
and three-dimensional systems than does the unbounded DOS of the hypercubic lattice.
Indeed, pathological results have been obtained on a hypercubic lattice due to the asymptotic freedom
of the quasiparticles in the tails of the Gaussian DOS \cite{che:03}.  As we shall see, the Bethe
lattice also has the advantage that analytic results are possible in the limit of small $\JH $,
precisely the regime where controversies persist.

The high-temperature, non-magnetic (NM) phases of the Heisenberg and DE
models have a correlation length $\xi $ that vanishes as $\zc \rightarrow \infty $.
By contrast, the short-range ordered (SRO) states introduced in an earlier
paper \cite{fis:06} possess some of the same characteristics as spin glasses:  local magnetic order and
exponentially decaying magnetic correlations \cite{bin:86}.  The SRO states are characterized by the
correlation parameter $q=\langle \sin^2 (\theta_i/2)\rangle $, where $\theta_i$ are
the angles between any spin and its neighbors.   Overall, the neighboring spins describe a cone
with angle $2\arcsin (\sqrt{q})$ around the central spin.  This disordered state reduces to the FM and
AF phases when $q=0$ and 1, respectively.   Unlike the FM and AF phases, a SRO state
with $0 < q < 1$ has extensive entropy and short-range but not long-range magnetic order.  Since it
is not possible to associate a wavevector with any SRO state, its transition temperatures $\TM (p,q)$
cannot be obtained from the magnetic susceptibility.  Rather, $\TM (p,q)$ can
be solved from coupled Green's function equations.  Remarkably, a SRO state can
be found with higher transition temperature and lower energy than the ordered FM and AF phases in a
large region of phase space.

The FM and AF transition temperatures may be obtained by solving either the coupled Green's function
equations or the Bethe-Salpeter (BS) equations for the magnetic susceptibility.  The BS equations break up into separate
relations for the charge and spin degrees of freedom.  The vertex functions of the BS equations
are consistent with the functional derivative of the self-energy with respect to the Green's function \cite{fis:05}.
It follows that the DMFT treatment of the DE model is $\Phi $-derivable and hence thermodynamically
consistent \cite{bay:62}, meaning that identical results are obtained from both ways of evaluating
$\TC $ and $\TN $.

This paper is divided into 7 main sections.  In Section II, we formulate the BS equations
for the susceptibility and derive the vertex functions.  The thermodynamic consistency of the
DMFT of the DE model is discussed in Section III.  The BS equations are used in
Section IV to derive implicit expressions for the FM and AF transition temperatures.
Using coupled Green's functions, we rederive the transition temperatures of the
ordered phases and evaluate the transition temperature of the SRO states.  We also explicitly solve those
relations in the weak-coupling limit of small $\JH $.  A complete map of the phase
instabilities of the DE model is provided in Section V.  In Section VI,
we present the $T=0$ phase diagram of the DE model both with and without the SRO states.
Finally, Section VII contains a summary and conclusion.  Appendix A provides results for
the transition temperatures and $T=0$ energies of the SRO states in the weak-coupling limit.
In Appendix B, we derive the self-consistency relation for the SRO states.
Appendix C provides a convenient and general form for the kinetic energy of the SRO and
ordered phases.

\section{Bethe-Salpeter Equations and the Vertex Functions}

The Hamiltonian of the DE model (also called the FM Kondo model when $\JH < \infty $) is
given by
\begin{equation}
\label{ham}
H=-t\sum_{\langle i,j \rangle }\Bigl( c^{\dagger }_{i\alpha }c^{\, }_{j\alpha }
+c^{\dagger }_{j\alpha }c^{\, }_{i\alpha } \Bigr) -2 \JH \sum_i \vs_i \cdot \vS_i
\end{equation}
where $c^{\dagger }_{i\alpha }$ and $c_{i\alpha }$ are the creation and destruction operators
for an electron with spin $\alpha $ at site $i$, $\JH >0 $ is the Hund's coupling,
$\vs_i = c^{\dagger }_{i\alpha } \bs_{\alpha \beta } c^{\, }_{i\beta }/2$
is the electronic spin, and $\vS_i=S\hm_i $ is the spin of the local moment.
Repeated spin indices are summed.  The local moment will be treated classically,
which is only a fair approximation for the magnanites with spin $S=3/2$ but a
better approximation for dilute magnetic semiconductors with $S=5/2$.
For large $\zc $, the hopping energy $t$ scales as $1/\sqrt{\zc }$
and the local effective action on site 0 may be written
\cite{fur:95}
\begin{equation}
\label{act}
\seff (\hm) =-T\sum_n\bc_{0\alpha }(i\nu_n) \Bigl\{ G_0(i\nu_n)^{-1}_{\alpha \beta }
+ \tJH \bs_{\alpha \beta }\cdot \hm\Bigr\} c_{0\beta }(i\nu_n),
\end{equation}
where $\tJH =\JH S$, $\nu_n=(2n+1)\pi T$, $\bc_{0\alpha }(i\nu_n)$ and $c_{0\alpha }(i\nu_n)$ are
now anticommuting Grassman variables, $\hm $ is the orientation of the local moment
on site 0, and $\uGr_0(i\nu_n)$ is the bare Green's function containing dynamical
information about the hopping of electrons from other sites onto site 0 and off again.

Because $\seff (\hm )$ is quadratic in the Grassman variables, the full Green's function
$\cG(\hm ,i\nu_n)_{\alpha \beta }$ at site 0 for a fixed $\hm $ may be readily solved by
integrating over the Grassman variables, with the NM result
\begin{equation}
\label{gg}
\cuG(\hm ,i\nu_n ) =\Bigl\{ G_0(i\nu_n)^{-1}\uI +\tJH \underline{\bs } \cdot \hm \Bigr\}^{-1}
=\ds\frac{G_0(i\nu_n)^{-1}\uI -\tJH \underline{\bs }\cdot \hm}{G_0(i\nu_n)^{-2}-\tJH^2 },
\end{equation}
where $G_0(i\nu_n)_{\alpha \beta }=\delta_{\alpha \beta } G_0(i\nu_n)$ is diagonal above $\TC $.
Averaging over the orientations of the local moment, $\langle \cG(\hm ,i\nu_n )_{\alpha \beta }\rangle_{\hm }
=G(i\nu_n)_{\alpha \beta }=\delta_{\alpha \beta }G(i\nu_n)$ where
\begin{equation}
\label{gga}
G(i\nu_n) =\ds\frac{G_0(i\nu_n)^{-1}}{G_0(i\nu_n)^{-2}-\tJH^2 }.
\end{equation}
If $P(\hm )= \Tr \Bigl( \exp (-\seff (\hm ))\Bigr) /Z$ is the
probability for the local moment to point in the $\hm $ direction,
then the average over $\hm $ is given by $\langle C(\hm
)\rangle_{\hm }\equiv \int d\Omega_{\hm } P(\hm )C(\hm )$. The
effective partition function $Z $ involves both a trace over the
Fermion degrees of freedom and an average over $\hm $: $Z =\int
d\Omega_{\hm } \Tr \Bigl( \exp (-\seff (\hm ))\Bigr) $. Above $\TC
$, $P(\hm )=1/4\pi $ is constant.  Consequently, the NM self-energy
is $\Sigma (i\nu_n ) = G_0(i\nu_n)^{-1}-G(i\nu_n)^{-1}=\tJH^2
G_0(i\nu_n)$.

Likewise above $\TC$, the lattice Green's function $\cG(\hm ,\vk ,i\nu_n)_{\alpha \beta }$ has the
expectation value
\begin{equation}
\label{lgf}
\langle \cG(\hm ,\vk ,i\nu_n )_{\alpha \beta }\rangle_{\hm }=\delta_{\alpha \beta }
\frac{1}{z_n-\epsilon_{\vk }-\tJH^2 G_0(i\nu_n)}
\end{equation}
with $z_n=i\nu_n +\mu $.
Summing Eq.(\ref{lgf}) over all $\vk $ and equating the result to Eq.(\ref{gga}), we
obtain \cite{fur:95,geo:96}
\begin{equation}
\label{sc}
G_0(i\nu_n)^{-1}=z_n  -\frac{W^2}{16}G(i\nu_n),
\end{equation}
where $W=4t\sqrt{\zc }$ is the full bandwidth of the semi-circular DOS.

\begin{figure}
\includegraphics *[scale=0.5]{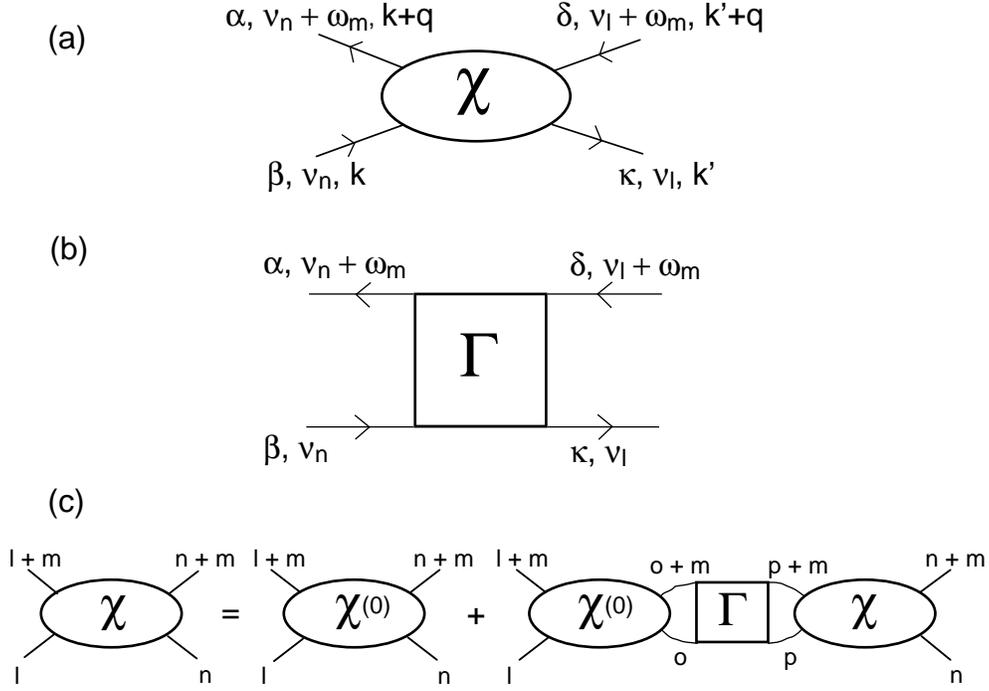}
\caption{
Diagrammatic representations of (a) the magnetic susceptibility $\chi^{\beta \alpha ,\delta \kappa }_{\vq ,i\om }(i\nu_l,i\nu_n)$,
(b) the vertex function $\Gamma_{i\om }^{\beta \alpha ,\delta \kappa }(i\nu_l,i\nu_n)$,
and (c) the Bethe-Salpeter equations.
}
\end{figure}

Our goal is to evaluate the general susceptibility \cite{fis:03}
\begin{equation}
\label{gsus}
\sum_{l,n} \chi^{\beta \alpha ,\delta \kappa }_{\vq ,i\om }(i\nu_l,i\nu_n)=
\int_0^{\beta }d\tau \, e^{i\om \tau }\sum_i
e^{-i\vq \cdot (\vR_i -\vR_0)}
\biggl\{
\langle T_{\tau } c_{i \alpha }^{\dagger }(\tau )c_{i \beta }(\tau ) c_{0 \kappa }^{\dagger } c_{0 \delta } \rangle
-\langle c_{i \alpha }^{\dagger }c_{i \beta }\rangle  \langle c_{0 \kappa }^{\dagger } c_{0 \delta } \rangle
\biggr\},
\end{equation}
where $\om =2m\pi T$ and $\langle C \rangle $ involves both a Fermion trace and an
average over all the $\hm_i$.  Sketched in Figs.1(a) and (b), the general susceptibility and the full vertex function
$\Gamma_{i\om }^{\eta \mu ,\tau \nu }(i\nu_l,i\nu_n)$ are related by the lattice BS equation:
\begin{equation}
\label{gBS}
\uc_{\vq ,i\om }^{\beta \alpha ,\delta \kappa }= \uc_{\vq ,i\om }^{(0)\, \beta \alpha ,\delta \kappa }
+\uc_{\vq ,i\om }^{(0)\, \beta \alpha ,\mu \eta }\uGm_{i\om }^{\eta \mu ,\tau \nu }
\uc_{\vq ,i\om }^{\nu \tau ,\delta \kappa }.
\end{equation}
Each term in this equation is a matrix in Matsubara space and matrix multiplication involves a
summation over intermediate Matsubara frequencies, as shown in Fig.2(c) \cite{fis:03}.
The local susceptibility $\uc_{i\om }^{\beta \alpha ,\delta \kappa }$ is obtained from
the lattice susceptibility $\uc_{\vq ,i\om }^{\beta \alpha ,\delta \kappa }$ by summing over $\vq $
and satisfies the local BS equation:
\begin{equation}
\label{glBS}
\uc_{i\om }^{\beta \alpha ,\delta \kappa }= \uc_{i\om }^{(0)\, \beta \alpha ,\delta \kappa }
+\uc_{i\om }^{(0)\, \beta \alpha ,\mu \eta }\uGm_{i\om }^{\eta \mu ,\tau \nu }
\uc_{i\om }^{\nu \tau ,\delta \kappa }.
\end{equation}
Since the vertex function is independent of momentum in infinite dimensions,
it appears identically in both the lattice and local BS equations \cite{jar:92,fre:95}.

For the bare susceptibilities, the electrons in each Green's function are uncorrelated,
which means that each Green's function is separately averaged over $\hm $.  So the bare,
lattice susceptibility is given by
\begin{equation}
\label{bsus}
\chi_{\vq ,i\om }^{(0)\, \beta \alpha ,\delta \kappa }(i\nu_l,i\nu_n )=-\frac{T}{N}\delta_{ln}
\sum_{\vk }\langle \cG(\hm ,\vk ,i\nu_l )_{\alpha \delta }\rangle_{\hm }
\langle \cG(\hm ,\vk +\vq,i\nu_{l+m})_{\kappa \beta }\rangle_{\hm }.
\end{equation}
The bare, local susceptibility is obtained by summing this expression over all $\vq $:
\begin{equation}
\label{blsus}
\chi_{i\om }^{(0)\, \beta \alpha ,\delta \kappa }(i\nu_l,i\nu_n )=-T\delta_{ln}
G(i\nu_l)_{\alpha \delta } G(i\nu_{l+m})_{\kappa \beta },
\end{equation}
where $G(i\nu_l)_{\alpha \beta }=\delta_{\alpha \beta }G(i\nu_l)$ is the $\hm $-averaged, local
Green's function given by Eq.(\ref{gga}).  It is also straightforward to evaluate the full,
local susceptibility:
\begin{eqnarray}
\label{lsus}
\chi_{i\om }^{\beta \alpha ,\delta \kappa }(i\nu_l,i\nu_n )&=
T\delta_{m,0}\Bigl\{ \langle \cG(\hm ,i\nu_n )_{\beta \alpha }\cG(\hm ,i\nu_l)_{\delta \kappa }\rangle_{\hm }
-\langle \cG(\hm ,i\nu_n)_{\beta \alpha }\rangle_{\hm }\langle \cG(\hm ,i\nu_l)_{\delta \kappa }\rangle_{\hm }
\Bigr\}\nonumber \\
&-T\delta_{ln}
\langle \cG(\hm ,i\nu_l)_{\alpha \delta } \cG(\hm ,i\nu_{l+m})_{\kappa \beta }\rangle_{\hm },
\end{eqnarray}
where $\cG(\hm ,i\nu_l)_{\alpha \beta }$ is the $\hm $-dependent, local Green's function given
by Eq.(\ref{gg}).

Due to the matrix structure of Eqs.(\ref{bsus}-\ref{lsus}),
the charge and spin degrees of freedom decouple:
$\sigma_z^{\alpha \beta }\uc_{\vq ,i\om }^{(0)\, \beta \alpha ,\delta \kappa } \propto \sigma_z^{\delta \kappa }$
and
$\delta^{\alpha \beta }\uc_{\vq ,i\om }^{(0)\, \beta \alpha ,\delta \kappa } \propto \delta^{\delta \kappa }$
with analogous relations for the local susceptibilities.
We define the charge and spin susceptibilities and vertex functions by
$\uc_{\vq ,i\om }^c=\delta_{\alpha \beta }\uc_{\vq ,i\om }^{\beta \alpha ,\delta \kappa }
\delta_{\kappa \delta }/4$,
$\uc_{\vq ,i\om }^{s}=\sigma^z_{\alpha \beta }\uc_{\vq ,i\om }^{\beta \alpha ,\delta \kappa }
\sigma^z_{\kappa \delta }/4$,
$\uGm_{i\om }^{c}=\delta_{\alpha \beta }\uGm_{\vq ,i\om }^{\beta \alpha ,\delta \kappa }
\delta_{\kappa \delta }$,
and
$\uGm_{i\om }^{s}=\sigma^z_{\alpha \beta }\uGm_{\vq ,i\om }^{\beta \alpha ,\delta \kappa }
\sigma^z_{\kappa \delta }$.
Similar definitions are employed for the local susceptibilities.
Since Eqs.(\ref{bsus}) and (\ref{blsus}) imply that the bare, charge and spin
susceptibilities are identical, we shall dispense with the superscripts $c$ and
$s$ on those quantities.

After this decoupling, both the lattice and local BS equations separate
into two sets of equations for the charge and spin degrees of freedom:
\begin{equation}
\label{BS}
\uc_{\vq ,i\om }^a= \uc_{\vq ,i\om }^{(0)}
+\uc_{\vq ,i\om }^{(0)}\uGm_{i\om }^a \uc_{\vq ,i\om }^a,
\end{equation}
\begin{equation}
\label{BSL}
\uc_{i\om }^a= \uc_{i\om }^{(0)}
+\uc_{i\om }^{(0)}\uGm_{i\om }^a \uc_{i\om }^a,
\end{equation}
where $a=c$ or $s$.  The same charge and spin vertex functions enter the lattice and local
equations.  The spin BS equations with $a=s$ were previously
solved \cite{fis:03} in the strong-coupling limit $\tJH \gg W$ and $\tJH \gg T$.

To generally solve Eq.(\ref{BS}) for the charge and spin lattice susceptibilities, we first
obtain the charge and spin vertex functions from the local susceptibilities given by Eq.(\ref{BSL}).
The vertex function may be written as
\begin{equation}
\label{gvf}
\Gamma_{i\om }^{\beta \alpha ,\delta \kappa }(i\nu_l,i\nu_n)=A_{ln}\bs_{\beta \alpha }\cdot
\bs_{\delta \kappa }\delta_{m,0}
+B_m^l\bs_{\delta \alpha }\cdot \bs_{\beta \kappa }\delta_{ln},
\end{equation}
\begin{equation}
A_{ln}=\frac{\beta \tJH^2}{3(1+C)}\frac{a_l a_n}
{b_l b_n},
\end{equation}
\begin{equation}
B_m^l=-\frac{\beta \tJH^2}{3}\frac{G_0(i\nu_l) G_0(i\nu_{l+m}) a_l a_{l+m}}
{G_0^{-1}(i\nu_l)G_0^{-1}(i\nu_{l+m})+\tJH^2},
\end{equation}
where $a_n=G_0(i\nu_n)^{-2}-\tJH^2$, $b_n=G_0(i\nu_n)^{-2}-\tJH^2/3$, and
$C=-(2\tJH^2/3)\sum_n b_n^{-1}$.
In terms of $A_{ln}$ and $B_m^l$, the charge and spin vertex functions are then given by
$\Gamma^c_{i\om }(i\nu_l,i\nu_n)=6B_m^l\delta_{ln}$ and
$\Gamma^s_{i\om }(i\nu_l,i\nu_n)=4A_{ln}\delta_{m,0}-2B_m^l\delta_{ln}$.

\section{Thermodynamic Consistency}

A thermodynamically consistent theory produces identical results from
either the Green's function (which is on the single-particle level) or the partition
function (which involves interactions on the two-particle level).
Baym and Kadanoff \cite{bay:62} demonstrated that a sufficient condition
for a theory to be thermodynamically consistent is that it be $\Phi $-derivable, which
means that a functional $\Phi (\{ \uGr (i\nu_n)\}) $, constructed from compact diagrams
involving the full Green's functions and the bare vertex functions, satisfies the condition
$\Sigma (i\nu_n)_{\alpha \beta }=\pt \Phi /\pt G(i\nu_n)_{\alpha \beta }$.
The partition function of a $\Phi $-derivable theory satisfies the relation
\begin{equation}
\label{z1}
-\log Z = \Phi -\sum_n \Tr \Bigl\{ \underline{\Sigma} (i\nu_n)\uGr (i\nu_n)\Bigr\}
+ \sum_n \Tr \log \Bigl\{ \uGr (i\nu_n)\Bigr\},
\end{equation}
which is stationary under variations of $\uGr (i\nu_n)$.  Whereas Baym and
Kadanoff considered systems of interacting Fermions and Bosons, their ideas were
later extended to systems of interacting electrons and spins \cite{bic:87} and to
disordered alloys \cite{lea:68}.  The $\Phi $-derivability of the DMFT of the
DE model was first demonstrated in Ref.\cite{fis:05}.

The bare vertex function $\Gamma_{i\om }^{(0)\, \beta \alpha ,\delta \kappa }(i\nu_l,i\nu_n)$
may be associated with the two-particle interaction in the purely electronic effective
action \cite{abr:63}
\begin{eqnarray}
\label{two}
&\seff' =-T\ds\sum_n\bc_{0\alpha }(i\nu_n) G_0(i\nu_n)^{-1} c_{0\alpha }(i\nu_n)\nonumber \\ &
-\ds\frac{T^3}{4}\ds\sum_{l,n,m}\bc_{0\alpha }(i\nu_n+i\omega_m)c_{0\beta }(i\nu_n)
\Gamma_{i\om }^{(0)\, \beta \alpha , \delta \kappa }(i\nu_l,i\nu_n)
\bc_{0\kappa }(i\nu_l)c_{0\delta }(i\nu_l+i\omega_m).
\end{eqnarray}
Because the Fermion operators anticommute, the bare vertex function must satisfy the crossing symmetries
$\Gamma_{i\omega_{n-l}}^{(0)\, \delta \alpha ,\beta \kappa }(i\nu_l,i\nu_{l+m})
=\Gamma_{i\omega_{l-n}}^{(0)\, \beta \kappa ,\delta \alpha }(i\nu_{n+m},i\nu_n)
=-\Gamma_{i\om }^{(0)\, \beta \alpha ,\delta \kappa }(i\nu_l,i\nu_n).$
There are two ways to evaluate the bare vertex function.
First, we can take the $\tJH \rightarrow 0$ limit of the full vertex function
$\uGm_{i\om }^{\beta \alpha , \delta \kappa }$ given by Eq.(\ref{gvf}).
Alternatively, we can associate the lowest-order, $\tJH^2$ contribution to the partition function
$Z =\int d\Omega_{\hm } \Tr \Bigl( \exp (-\seff (\hm )) \Bigr) $
with the $\tJH^2$ contribution to the partition function $Z' =\Tr \Bigl( \exp (-\seff')\Bigr)$.
Both methods yield the same result:
\begin{equation}
\label{G0}
\uGm_{i\om }^{(0)\, \beta \alpha ,\delta \kappa }(i\nu_l,i\nu_n)
=\frac{1}{3}\beta \tJH^2 \Bigl\{
\bs_{\beta \alpha }\cdot \bs_{\delta \kappa }\delta_{m,0}-\bs_{\delta \alpha }\cdot \bs_{\beta \kappa }
\delta_{ln}\Bigr\},
\end{equation}
which does indeed satisfy the required crossing symmetries.

But $\seff'$ produces an inequivalent infinite-dimensional theory than produced by $\seff (\hm )$,
as seen in the different expansions of $Z$ and $Z'$ to order $\tJH^4$:
\begin{equation}
Z=Z_0\Biggl\{ 1 -\tJH^2 \sum_n G_0(i\nu_n)^2 +\frac{1}{2}\tJH^4 \sum_{l \ne n}G_0(i\nu_l)^2G_0(i\nu_n)^2+{\cal O}(\tJH^6) \Biggr\} ,
\end{equation}
\begin{equation}
Z'=Z_0\Biggl\{ 1 -\tJH^2 \sum_n G_0(i\nu_n)^2 +\frac{5}{6}\tJH^4 \sum_{l \ne n}G_0(i\nu_l)^2G_0(i\nu_n)^2+{\cal O}(\tJH^6) \Biggr\}.
\end{equation}
Therefore, the DE action cannot be replaced by a purely electronic action containing only
two-body interactions and it is not possible to construct a diagrammatic expansion
in powers of $\an $.

Nevertheless, we can construct a functional $\Phi (\{ \uGr (i\nu_n)\})$ satisfying
the condition
$\Sigma (i\nu_n)_{\alpha \alpha }= \pt \Phi /\pt G(i\nu_n)_{\alpha \alpha }$.
Starting from Eq.(\ref{gg}) and Dyson's equation for the self-energy, we obtain
$\pt \Sigma (i\nu_l)_{\alpha \alpha }/\pt G(i\nu_n)_{\beta \beta }=(\uK^{-1})_{ln}^{\alpha \beta }
+\delta_{ln}\delta_{\alpha \beta }G(i\nu_n)^{-2}$, where
\begin{equation}
K^{\alpha \beta }_{ln}= \ds\frac{\pt G(i\nu_n)_{\beta \beta }}{\pt [G_0(i\nu_l)_{\alpha \alpha }]^{-1}}
=-\delta_{ln}\ds\frac{1}{a_n^2}\Biggl\{ \ds\frac{2\tJH^2}{3}+b_n \delta_{\alpha \beta }\Biggr\}
+\ds\frac{\tJH^2}{3a_la_n}\Bigl(2\delta_{\alpha \beta }-1\Bigr).
\end{equation}
Inverting this Jacobian yields the general result
\begin{eqnarray}
\label{jac}
\ds\frac{\pt \Sigma (i\nu_l)_{\alpha \alpha }}{\pt G(i\nu_n)_{\beta \beta }}
&=&-\delta_{ln}\ds\frac{\tJH^2 a_n^2}{3b_n}\Biggl\{ \ds\frac{2}{2a_n-3b_n}
+\delta_{\alpha \beta }G_0(i\nu_n)^2\Biggr\}
-\ds\frac{\tJH^2 }{3(1+C)}\ds\frac{a_la_n}{b_lb_n}
\Bigl( 2\delta_{\alpha \beta }-1\Bigr)\nonumber \\
&=&-T\,\Gamma_{i\om =0}^{\alpha \alpha ,\beta \beta }(i\nu_l,i\nu_n),
\end{eqnarray}
where the right-hand side is the full vertex function of Eq.(\ref{gvf}).
The functional $\Phi $ must exist because the vertex function is well-defined and
the curl of the self-energy vanishes:
$\pt \Sigma (i\nu_n)_{\alpha \alpha }/\pt G(i\nu_l)_{\beta \beta  } -
\pt \Sigma (i\nu_l)_{\beta \beta }/\pt G(i\nu_n)_{\alpha \alpha } =0$.
So despite the absence of a diagrammatic perturbation theory,
the DMFT of the DE model is $\Phi $-derivable and hence, thermodynamically
consistent.

\section{Transition Temperatures}

Since the electronic spin-susceptibility is given by
$\cel (\vq )=\sum_{l,n}\chi^s_{\vq ,i\om =0}(i\nu_l,i\nu_n) $,
the condition that $\cel (\vq )$ diverges is equivalent to the condition
that $\Det (\uc^s_{\vq, i\om=0})^{-1} =0$.  To obtain the FM or AF
transition temperatures, we require the $\vq =0$ or $\vQ $ susceptibilities with $\om =0$.
The AF wavevector $\vQ $ is defined so that the Fermi surface is nested at half
filling or $\epsilon_{\vk +\vQ}=-\epsilon_{\vk }$.
Although the Bethe lattice sketched in Fig.1 lacks translational symmetry and nonzero wavevectors
$\vq $ are ill-defined, the AF wavevector $\vQ $ may be associated with the bipartite nature of the Bethe lattice.
To be safe, the N\'eel temperature will be evaluated in two different ways, only one of which relies on
the existence of an AF wavevector.

For either FM or AF ordering, the electron filling $p$ can be related to the
chemical potential $\mu $ through the relation
\begin{equation}
\label{pl}
p=1+T\sum_n \re G(i\nu_n) =1-\frac{32}{W^2}T\sum_n\re R_n,
\end{equation}
where $G_0(i\nu_n)^{-1}= z_n+R_n$ and the complex function $R_n$ satisfies the
implicit relation
\begin{equation}
\label{Rl}
a_n R_n + \frac{W^2}{16}(z_n+R_n)=0.
\end{equation}
Since $a_n=(z_n+R_n)^2-\tJH^2$, $R_n$ falls off like $-W^2/(16z_n)$ for large $n$.
Notice that $p=1$ when $\mu =0$.  Since all of our results are particle-hole symmetric about $p=1$, we shall
restrict consideration to the range $0 \le p\le 1$.

As discussed in Section II, the BS equations decouple into charge and spin relations.
Since the charge vertex function $\Gamma^c_{i\om }(i\nu_l,i\nu_n)=6B_m^l\delta_{ln}$ is
diagonal in Matsubara space, the condition $\Det (\uc^c_{\vq, i\om=0})^{-1} =0$ has no solution
for $\vq =0 $ or $\vQ$.  In other words, there is no charge-density wave instability within the DE model.

\subsection{Ferromagnetic ordering}

When $\vq =0$, the bare, zero-frequency lattice susceptibility is given by
\begin{equation}
\chi^{(0)}_{\vq =0,i\om =0}(i\nu_l,i\nu_n)=\frac{4T}{W^2}\Biggl( 1-\frac{z_l-\tJH^2 G_0(i\nu_l)}{\sqrt{
(z_l-\tJH^2 G_0(i\nu_l))^2-W^2/4}}\Biggr)\delta_{ln}.
\end{equation}
With the help of Eq.(\ref{Rl}), we find that the full spin-susceptibility at $\vq =0$ satisfies
\begin{eqnarray}
&\chi^{s\, -1}_{\vq =0,i\om =0}(i\nu_l,i\nu_n) = \chi^{(0) -1}_{\vq =0,i\om =0}(i\nu_l,i\nu_n)-\Gamma^s_{i\om =0}(i\nu_l,i\nu_n)
\nonumber \\
&=-2\beta \delta_{ln}\ds\frac{a_l}{(z_l+R_l)b_l}\Bigl\{ (z_l+R_l)^2(z_l+2R_l)-\tJH^2(z_l+4R_l/3)\Bigr\}
-\ds\frac{4\beta \tJH^2}{3(1+C)}\ds\frac{a_l a_n}{b_l b_n},
\end{eqnarray}
where $b_l =(z_l+R_l)^2-\tJH^2/3$.  Consequently, the condition that
$\Det (\uc^s_{\vq =0, i\om=0})^{-1} =0$ can be written
as an implicit condition for the Curie temperature:
\begin{equation}
\label{TCF}
\frac{2\tJH^2}{3}\sum_n \frac{R_n}{(z_n+R_n)^2(z_n+2R_n)-\tJH^2(z_n+4R_n/3)}=1,
\end{equation}
which must be supplemented by Eq.(\ref{pl}) for the electron filling.

If $\tJH /W > 1/4$, the interacting DOS splits into two bands \cite{che:03}.
In the lower band centered at $-\tJH $ with $0 \le p \le 1$, all electrons have their spins parallel to
the local moments;  in the upper band centered at $\tJH $ with $1 \le p \le 2$, all holes
have their spins parallel to the local moments.  As $\tJH /W $ increases, both upper
and lower bands are narrowed by correlations and their widths approach $W'=W/\sqrt{2}$.
To recover the large $\tJH $ limit, the chemical potential must be written as
$\mu =\tJH \, \sgn (p-1) +\delta \mu $ where $-W'/2 \le \delta \mu \le W'/2$.
Eq.(\ref{TCF}) then reduces to \cite{aus:01,fis:03}
\begin{equation}
\label{TCin}
\sum_n \frac{R_n^2}{R_n^2-3W^2/32}=1,
\end{equation}
where $R_n$ is now given explicitly by $R_n= \Bigl(\sqrt{z_n^2-W^2/8}-z_n\Bigr)/2$.
So in this limit, $\TC $ is proportional to $W$ and is independent of $\tJH $.
Due to the strong Hund's coupling, $\TC $ vanishes at half filling ($p=1$) or when the
lower band is completely full because electrons are unable to hop to neighboring sites.

In the weak-coupling regime of $T \ll \tJH \ll W$, the summation in Eq.(\ref{TCF})
may be replaced by an integral over frequency $\nu $, as shown in Appendix A.
Carefully treating the branch cut along the real axis between $\pm W/2$, we find that
\begin{equation}
\label{TCwc}
\TC =\frac{32\tJH^2}{9\pi W}\sqrt{1-(2\mu /W)^2}\Bigl\{ (2\mu /W)^2-1/4 \Bigr\},
\end{equation}
which is proportional to $\tJH^2/W$ as expected in the RKKY limit.
For small filling ($\mu $ close to $-W/2$) and a fixed $\tJH $, $\TC $ initially
increases with $p$ and then begins to decrease, vanishing at $\mu =W/4$ corresponding
to a filling of $p\approx 0.39$.  Because $\TC (p,\tJH )$ is particle-hole symmetric,
it is the same for $p$ electrons ($2-p$ holes) and $2-p$ electrons ($p$ holes)
per site.  These results are in qualitative agreement with Nolting {\em et al.} \cite{nol:97,san:02},
who evaluated the Curie temperature of the $s-f$ model with quantum local spins
using both equations-of-motion \cite{nol:97} and modified RKKY \cite{san:02} techniques.
Nolting {\em et al.} obtained FM solutions for small $\tJH$ in a concentration range
significantly narrower than found here, probably because their approach retains the non-local
fluctuations produced by the RKKY interaction.

For small fillings and $\varepsilon \equiv 2\mu /W +1 \ll 1$, $\TC W/\tJH^2\approx 8\sqrt{2}\varepsilon^{1/2}/(3\pi )$.
Since $p\approx 8\sqrt{2}\varepsilon^{3/2}/(3\pi )$, $\TC $ grows like $p^{1/3}$ for small $p$.

\subsection{Antiferromagnetic ordering}

This section presents two calculations for the AF N\'eel temperature:
one based on the magnetic susceptibility and the other on coupled Green's functions.  Their
equivalence was established in Section III, which demonstrated that the DMFT
of the DE model is thermodynamically consistent.  While the first technique explicitly
assumes the existence of an AF wavevector $\vQ$, the second does not.

\subsubsection{From the susceptibility}

Evaluating the N\'eel temperature from the susceptibility requires the bare, zero-frequency lattice
susceptibility for $\vq = \vQ $:
\begin{equation}
\chi^{(0)}_{\vQ ,i\om =0}(i\nu_l,i\nu_n)=-\frac{4T}{W^2}
\Biggl( 1-\frac{\sqrt{(z_l-\tJH^2 G_0(i\nu_l))^2-W^2/4}}
{z_l-\tJH^2 G_0(i\nu_l)}
\Biggr)\delta_{ln}.
\end{equation}
After some algebra, we find that the full spin-susceptibility at $\vq =\vQ$ satisfies
\begin{eqnarray}
&\chi^{s\, -1}_{\vQ ,i\om =0}(i\nu_l,i\nu_n) = \chi^{(0) -1}_{\vQ ,i\om =0}(i\nu_l,i\nu_n)-\Gamma^s_{i\om =0}(i\nu_l,i\nu_n)
\nonumber \\
&=-2\beta \delta_{ln}\ds\frac{a_l}{(z_l+R_l)b_l}\Bigl\{ z_l(z_l+R_l)^2-\tJH^2(z_l+2R_l/3)\Bigr\}
-\ds\frac{4\beta \tJH^2}{3(1+C)}\ds\frac{a_l a_n}{b_l b_n}.
\end{eqnarray}
Consequently, $\TN $ is found from the condition that $\Det (\uc^s_{\vQ , i\om=0})^{-1} =0$
or
\begin{equation}
\label{TCN}
-\frac{2\tJH^2}{3}\sum_n \frac{R_n}{z_n(z_n+R_n)^2-\tJH^2(z_n+2R_n/3)}=1,
\end{equation}
which bares some similarity to Eq.(\ref{TCF}) for $\TC $.

In the strong-coupling limit, Eq.(\ref{TCN}) reduces to
\begin{equation}
\label{TNin}
\sum_n \frac{R_n^2}{R_n^2+3W^2/32}=1,
\end{equation}
with a plus sign multiplying $3W^2/32$ compared to the minus sign in Eq.(\ref{TCin}).  It is
straightforward to show that $\TN \le 0$ for all $p$ in this limit.  Hence, the AF
phase instabilities disappear as $\tJH /W \rightarrow \infty $.

In the weak-coupling regime, we may once again replace the Matsubara sum by an integral,
with the result
\begin{equation}
\label{TNwc}
\TN =\frac{8\tJH^2}{3\pi W}\Biggl\{
\log \Biggl( \frac{1+\sqrt{1-(2\mu /W)^2}}{2\vert \mu \vert /W}\Biggr) -
\sqrt{1-(2\mu /W)^2} -\frac{4}{3}\Bigl( 1-(2\mu /W)^2\Bigr)^{3/2}\Biggr\}.
\end{equation}
For $p < 0.73$, $\TN $ is negative so the system becomes FM for $0< p \le 0.39$.

Close to half filling, $\TN $ given by Eq.(\ref{TNwc}) becomes positive and the DE model supports an
AF phase.  We find that $\TN > 0$ for $0.73 \le p \le 1$, in which range $\TC $
given by Eq.(\ref{TCwc}) is negative.  The divergence of $\TN $ in Eq.(\ref{TNwc}) as $p\rightarrow 1$
or $\mu \rightarrow 0$ signals a breakdown in the weak-coupling expansion and the appearance of
a gap in the interacting DOS, as discussed later.

\subsubsection{From coupled Green's functions}

The previous sub-section used the nesting condition for the
AF wavevector $\vQ$ whereas, strictly speaking, nonzero wavevectors are not
defined on the Bethe lattice.  Hence, it is worthwhile to re-evaluate $\TN $ using
a Green's function approach that skirts the question of whether an AF wavevector exists.
This Green's function approach will also come in handy when considering the SRO state
later in Section IV.

Dividing the Bethe lattice into $A$ and $B$ sub-lattices, we obtain coupled relations between
the bare and full local Green's functions:
\begin{equation}
\label{scsl}
\uGr_0^{(\al )}(i\nu_n)^{ -1}=z_n \uI -\frac{W^2}{16}\uGr^{(\bal )}(i\nu_n),
\end{equation}
where $\al = \uparrow $ or $\downarrow $ on the $A$ or $B$ sites and $\bal $ is opposite to $\al $.
If the bare inverse Green's function is parameterized as
\begin{equation}
\label{g0}
\uGr_0^{(\al )}(i\nu_n )^{-1}=(z_n+R_n)\uI +Q_n^{(\al )}\underline{\sigma }_z,
\end{equation}
then Eq.(\ref{scsl}) implies that $R_n$ and $Q_n^{(\al )}$ are solved
from the condition
\begin{equation}
R_n\uI +Q_n^{(\al )}\underline{\sigma }_z =-\frac{W^2}{16} \Bigl\langle \Bigl\{ (z_n+R_n)\uI +
\underline{\bs }\cdot \Bigl( \tJH \hm +Q_n^{(\bal )}\bz \Bigr)\Bigr\}^{-1} \Bigr\rangle_{\bal }.
\end{equation}
To linear order in the sublattice magnetization, $R_n$ and $Q_n^{(\al )}$ satisfy the implicit relations
\begin{equation}
\label{Rn}
a_nR_n=-\frac{W^2}{16}(z_n+R_n),
\end{equation}
\begin{equation}
Q_n^{(\al )}=\frac{R_n}{z_n+R_n}\Biggl\{ Q_n^{(\bal )}\Biggl( 1+\frac{2\tJH^2 }{3a_n}\Biggr)  +\tJH M^{(\bal )}
\Biggr\},
\end{equation}
where $ \langle \hm \rangle_{\al }\equiv M^{(\al )}\bz =\pm M\bz $, the sign depending on whether $\al =\uparrow $ (A site)
or $\al =\downarrow $ (B site).

The linear relations for $Q_n^{(\al )}$ may be readily solved, with the result
\begin{equation}
\label{Qn}
Q_n^{(\al )}=\frac{M^{(\al )}\tJH a_n R_n}{(z_n+R_n)^2z_n-\tJH^2 (z_n+2R_n/3)}.
\end{equation}
After integrating $\exp (-\seff (\hm ))$ over the Grassman variables, we find that the probability
for the local moment on sublattice $\al $ to point along $\hm $ is
\begin{equation}
P(\hm )^{(\al )} \propto \exp \Biggl\{ \sum_n \log \Biggl( 1 - \frac{2\tJH Q_n^{(\al )} m_z}{a_n }\Biggr)
\Biggr\} \propto \exp (\beta J_{eff} M^{(\al )} m_z ).
\end{equation}
The last relation defines the effective interaction:
\begin{equation}
\label{jeff}
J_{eff} =-\ds\frac{2T\tJH }{M^{(\al )}}\sum_n \ds\frac{Q_n^{(\al )}}{a_n}
=-2T\tJH^2 \sum_n \ds\frac{R_n}{z_n(z_n+R_n)^2-\tJH^2 (z_n+2R_n/3)},
\end{equation}
which is the same on each sublattice.
Finally, $\TN $ is solved from the condition $M^{(\al )}=J_{eff}M^{(\al )}\beta /3$,
which reproduces Eq.({\ref{TCN}}).  So the earlier results for $\TN $ do not
depend on the existence of an AF wavevector!

\subsection{Other magnetic phases}

Earlier studies of the Heisenberg model on the Bethe lattice \cite{tri:83,rog:97} proposed
a spiral magnetic state with a wavevector that characterizes the twist of the spin between
nearest neighbors.  But it is quite easy to show that a Bethe lattice can not be
partitioned into more than two sublattices.  For example, if we try to partition a
Bethe lattice with $\zc =4$ into three sublattices, then sites $C$ and $A$ (red and purple dots
in Fig.3(a)) would be next-nearest neighbors but the two sites $C$ (purple dots)
(presumably on the same magnetic sublattice) are actually next-nearest neighbors
of each other.  So unless the central site is singled out, a Bethe lattice
does not support a magnetic state with more than two sublattices.

More recently, Gruber {\em et al.} \cite{gru:00} argued that this restriction may be
removed as $\zc \rightarrow \infty $ because the paths with the largest weight always move
outward from the central site.  This analysis uses the fact that the surface of the Bethe lattice
contains a non-vanishing fraction of its sites \cite{egg:74}.  Since the period-three state constructed
in Ref.\cite{gru:00} can only be formed around a central site, it is not isotropic.  Such states
will not be considered further in this paper.

\begin{figure}
\includegraphics *[scale=0.8]{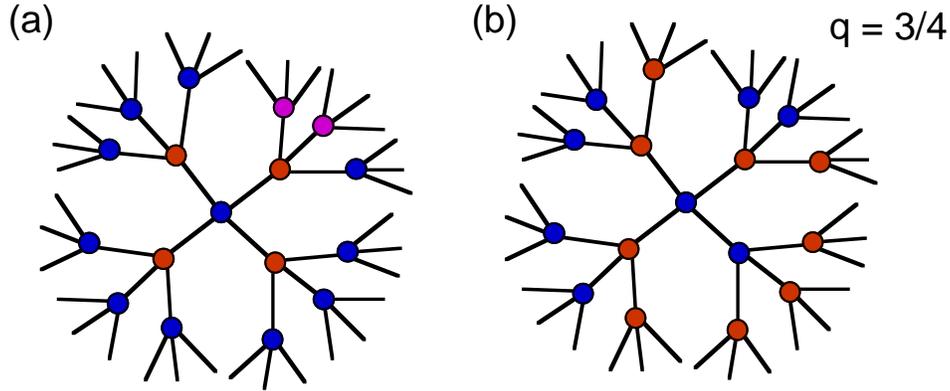}
\caption{
(a) A possible partitioning of the Bethe lattice into three sublattices with
blue, red, and purple sites.  (b) A colinear SRO state on a $\zc =4$ Bethe lattice with $q=3/4$,
meaning that 1/4 of the neighbors of any site have the same spin while 3/4 have the
opposite spin.  The two spins are denoted by blue and red dots.
}
\end{figure}

The broad class of SRO states first introduced in an earlier paper \cite{fis:06} is characterized by the
correlation parameter $q=\langle \sin^2 (\theta_i/2)\rangle $, where the average is
taken over all neighbors of any site on the Bethe lattice.  Since the azimuthal angles $\phi_i$ of the
neighboring spins are assumed to be random with $\langle \exp (i\phi_i)\rangle =0$, the neighboring
spins will on average sweep out a cone with angle $2\arcsin (\sqrt{q})$ about the central spin.
Due to the topology of the Bethe lattice, a SRO phase cannot be ordered (except in the FM or
AF limits when $q=0$ or 1).  For the special case of a colinear SRO state, $\theta_i=0$ or $\pi $
on every site and $q$ is the fraction of neighbors of any site that have the opposite spin.
An example of a colinear SRO state with $\zc =4$ and $q=3/4$ is sketched in Fig.3(b).

Quite generally, the SRO phase has short-range but not long-range magnetic order.
If $M_n$ is the average magnetization of the lattice sites at a distance $na$ (lattice constant
$a$) from a central site, then it is easy to show that $M_n=(1-2q)M_{n-1}$.
So the magnetization decays exponentially like $\vert M_n \vert =\vert M_0 \vert \exp(-na/\xi )$
with the correlation length $\xi = -a/\log \vert 2q-1\vert \ge 0$.  The
correlation length diverges in the FM and AF limits and vanishes in the NM state with
$q=1/2$, when the spins around any site are completely disordered.
A SRO phase can have either FM ($0 < q < 1/2$) or AF ($1/2 < q < 1$) correlations, as discussed further below.

Also working on the Bethe lattice, Chattopadhyay {\em et al.} \cite{cha:00} assumed that
all of the angles $\theta_i$ were identical and characterized the resulting state by its
``incommensurate spin correlations.''   We avoid this description
because nonzero wavevectors different are ill-defined on the Bethe lattice
and the $X({\bf q})$ formalism of M\"uller-Hartmann \cite{mul:89} cannot be adapted
from the hypercubic to the Bethe lattice.  In our view, the SRO phase is more accurately
characterized by the magnetic correlations around any site.

In the absence of the SRO states, a large portion of the magnetic phase diagram is NM.
This poses a problem because Nernst's theorem requires that the entropy vanishes
at $T=0$.   Generally, the entropy of a SRO state is difficult to evaluate.  But
in the $\zc \rightarrow \infty $ limit, the entropy of a colinear SRO phase  is
\begin{equation}
\frac{S_{{\rm col}}}{N}=-q\log q -(1-q)\log (1-q) ,
\end{equation}
which vanishes only in the FM or AF phases.  Of course, the entropy of a non-colinear
SRO phase must be greater than $S_{{\rm col}}$.  Therefore, the formal dilemma
posed by Nernst's theorem is not resolved by the presence of SRO states.
As discussed in the conclusion, entropic ground states frequently arise
in infinite dimensions.  In finite dimensions, the SRO state may evolve
into a state with incommensurate SRO.

Since a SRO phase with correlation parameter $q$ between 0 and 1 has no long-range order, the transition
temperature $\TM (p,q)$ cannot be obtained from the condition that the magnetic susceptibility
diverges.  But we can easily solve for $\TM (p,q)$ by generalizing the Green's function technique
used in sub-section IV.B.2.  With the SRO state defined above, the self-consistency relation between
the bare and full local Green's functions is derived in Appendix B:
\begin{equation}
\label{scsg}
\uGr_0^{(\al )}(i\nu_n)^{-1}=z_n \uI -\frac{W^2}{16}\Bigl\{
q\, \uGr^{(\bal )}(i\nu_n) +(1-q)\, \uGr^{(\al )}(i\nu_n)\Bigr\},
\end{equation}
where $\uGr^{(\al )}$ and $\uGr^{(\bal )}$ are spin-reversed Green's functions.  The quantization axis
$\al $ is defined by the spin on the central site 0.  The above relation was previously obtained in Ref.\cite{cha:00}
under the more restrictive condition that $\theta_i$  is constant.  Once again parameterizing the bare inverse Green's function
by Eq.(\ref{g0}), we find that $R_n$ is independent of the magnetization $M=\langle m_z\rangle $ on site $0$
(assumed spin up) and is given by Eq.(\ref{Rn}) while $Q_n^{(\uparrow )}=-Q_n^{(\downarrow )}$
has the linearized solution
\begin{equation}
Q_n^{(\uparrow )}=\frac{(2q-1)M\tJH a_n R_n}{(z_n+R_n)^2(z_n+2(1-q)R_n)-\tJH^2
(z_n+2(2-q)R_n/3)},
\end{equation}
which reduces to Eq.(\ref{Qn}) when $q=1$.

The effective interaction $J_{eff}$ is again given by the first equality in Eq.(\ref{jeff}) so that
$\TM (p,q)$ is solved from the implicit relation
\begin{equation}
-\frac{2\tJH^2}{3}(2q-1)\sum_n \frac{R_n}{(z_n+2(1-q)R_n)(z_n+R_n)^2-\tJH^2(z_n+2(2-q)R_n/3)}=1,
\end{equation}
which reduces to both the FM and AF results, Eq.(\ref{TCF}) and (\ref{TCN}), when
$q=0$ and $q=1$.    It also follows that $\TM (p,q)=0$ in the NM state with $q=1/2$.

\begin{figure}
\includegraphics *[scale=0.6]{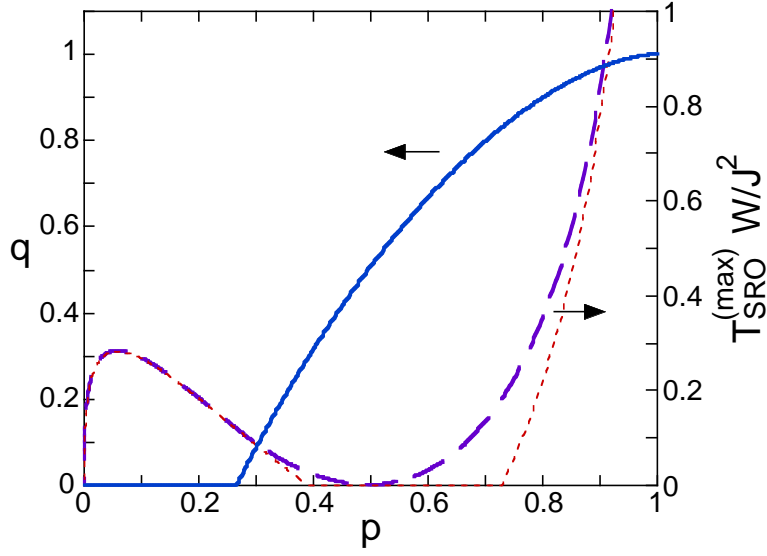}
\caption{
The correlation parameter $q$ (solid) and the associated maximum SRO transition temperature
$\TR (p)$ (dashed) versus filling $p$
for the SRO states in the weak-coupling limit.  Also plotted in the thin short-dashed line are
the weak-coupling limits of $\TC $ for $q=0$ and $\TN $ for $q=1$.
}
\end{figure}

In the strong-coupling limit $W \ll \tJH $ and $T\ll \tJH$, $\TM (p,q) $ is solved from the condition
\begin{equation}
\sum_n \frac{R_n^2}{R_n^2+3W^2/(32(2q-1))}=1.
\end{equation}
It is straightforward to show that $\TM (p,q)$ is negative for
$1/2\le q \le 1$ and is positive but always smaller than $\TC (p)$
for $0\le q < 1/2$.  So as expected, the SRO states are unstable in the
strong-coupling limit.

Results for $\TM (p,q)$ in the weak-coupling limit are given in Appendix A.  Maximizing
$\TM (p,q)$ with respect to $q$, we obtain $\TR (p)$ plotted
in Fig.4.  Remarkably, a SRO phase with AF correlations can always be found with
a higher transition temperature than the AF state.  As $q\rightarrow 1$, the AF and SRO
transition temperatures meet at $p=1$ or half filling.  Even in the range of fillings
between about 0.26 and 0.39, where $\TC $ is nonzero, a SRO phase with
FM correlations or $0 < q < 1/2$ has the higher critical temperature!  Also notice that the SRO transition
temperature is nonzero in the range of fillings between 0.39 and 0.73, where neither the FM nor
AF states was stable for small $\tJH /W$.  The point at which $q=1/2$ and $\TR (p)=0$
lies slightly below $p=0.5$.   Recall that  $q$ changes discontinuously at $\TR (p)$ from
1/2 in the NM phase above to a value greater than or smaller than 1/2 in the SRO phase below.

\section{Magnetic Instabilities of the Double-Exchange Model}

We shall first present results for the magnetic transition temperatures
in the absence of the SRO states.  These results are then revised to include the
SRO states.  Bare in mind that just because a FM, AF, or SRO state has
the highest transition temperature for a given $\tJH /W$ and $p$
does not imply that it remains stable down to zero temperature or
even that it is reached at all:  a pure state may be bypassed by
the formation of a PS mixture.  The issue of PS can only be resolved by
calculating the free energy as a function of filling.  We shall perform
such a study for $T=0$ in Section VI.

\begin{figure}
\includegraphics *[scale=0.4]{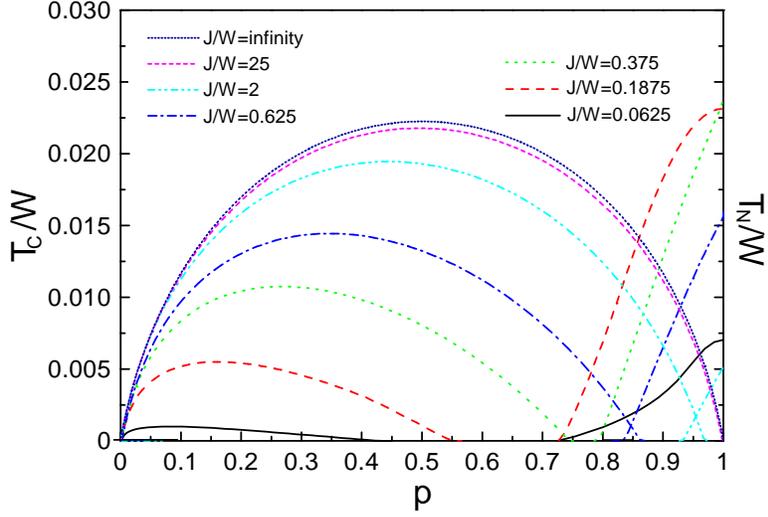}
\caption{
The Curie and N\'eel temperatures versus filling $p$ for a variety of couplings
$\tJH /W$.
}
\end{figure}

In Fig.5, the Curie and N\'eel temperatures are plotted versus filling for a variety
of couplings.  When $p \ge 0.39$, a critical value of $\tJH $ is required
for $\TC $ to be nonzero.  So the weak-coupling regime is restricted to
electron concentrations smaller than 0.39.
As $\tJH /W$ increases, the AF instabilities are restricted to a narrower
range of concentrations near $p=1$.  For large $\tJH /W$ at half filling, the
nearest-neighbor hopping can be treated perturbatively and superexchange
leads to a N\'eel temperature proportional to $W^2/\tJH $.  So the sliver of AF phase stabilized
around half filling rapidly narrows with increasing $\tJH /W$.  By contrast the FM phase
dominates as $\tJH /W$ increases.  In the limit $\tJH /W\rightarrow \infty $, the N\'eel temperature
vanishes even at $p=1$.

\begin{figure}
\includegraphics *[scale=0.4]{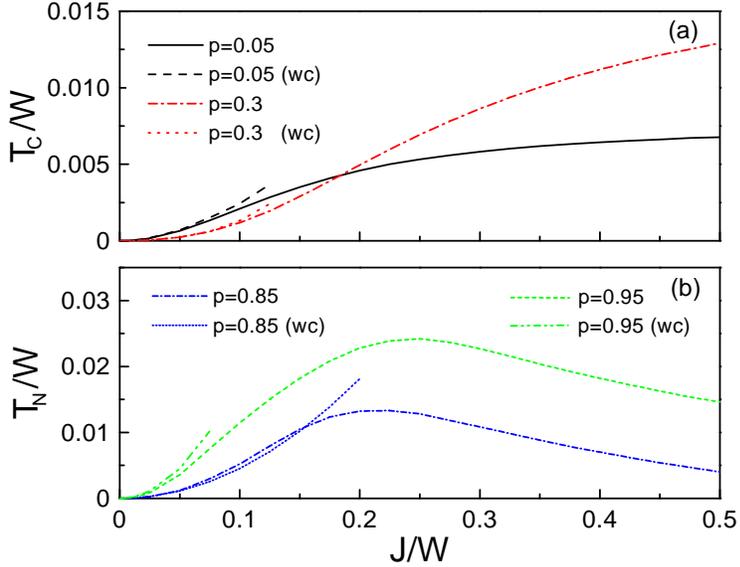}
\caption{
The (a) Curie and (b) N\'eel temperatures versus $\tJH /W $ for several fillings.  Also
shown are the weak-coupling transition temperatures obtained using perturbation
theory.
}
\end{figure}

For a fixed filling, $\TC $ and $\TN $ are plotted versus $\tJH /W$ in Fig.6.
As expected, the weak-coupling curves track the numeric results for small
$\tJH /W$ but deviate for larger $\tJH /W$.
Whereas the Curie temperature reaches a plateau at a value of order $W$,
the N\'eel temperature for any $p<1$ vanishes at a finite value of $\tJH /W$.
Again, the filling range of AF instabilities rapidly narrows as $\tJH /W$ increases.

\begin{figure}
\includegraphics *[scale=0.6]{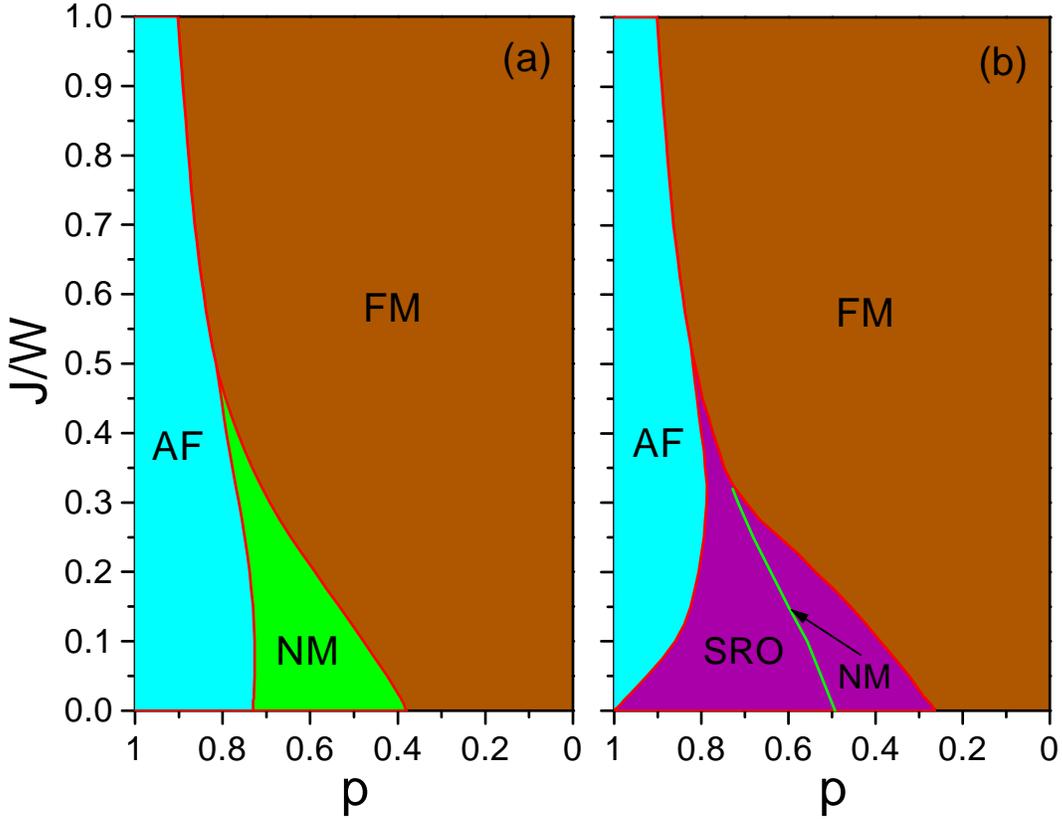}
\caption{
The phase instabilities of the DE model (a) assuming only FM, AF, and NM phases
and (b) also including the SRO phase.
}
\end{figure}

The phase instabilities of the DE model in the absence of SRO states are plotted in Fig.7(a).
The NM region shrinks as $\tJH /W$ increases and is absent
for $\tJH /W > 0.5$.  Since the interacting DOS splits for $\tJH /W > 1/4$ \cite{che:03},
the NM region survives even after the formation of sub-bands.  Whereas the AF region persists
at half filling for all $\tJH /W$, the sliver of AF phase rapidly narrows with increasing coupling.
The FM/NM phase boundary in Fig.7(a) agrees quite well with that found by Auslender {\em et al.} \cite{aus:01}.

\begin{figure}
\includegraphics [scale=0.5]{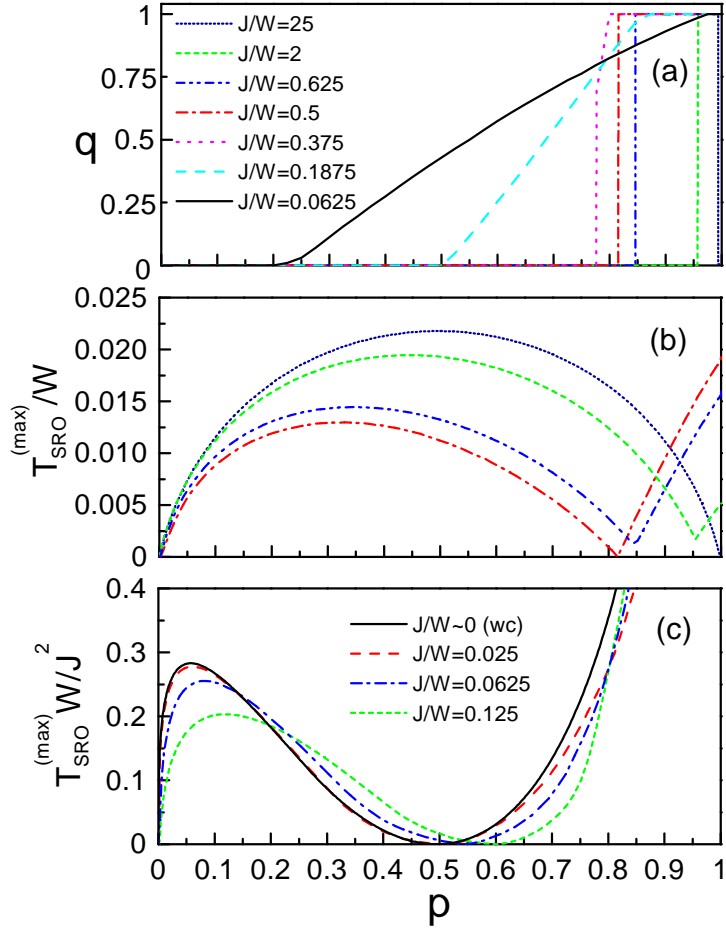}
\caption{
(a) The correlation parameter $q$, (b) the maximum transition temperature $\TR (p)/W$,
and (c) the normalized transition temperature $\TR (p)W/\tJH^2$ for various
values of $\tJH /W$ versus filling $p$.
}
\end{figure}

Now including the SRO states, we plot the correlation parameter $q$ along with $\TR (p)$ in Fig.8.
For $\tJH /W >0$, a transition into the AF state ($q=1$) occurs in a narrow range of
fillings close to $p=1$.  This range vanishes both as $\tJH /W \rightarrow 0$ and as
$\tJH /W \rightarrow \infty $.  It has a maximal extent (from $p=0.78$ to 1) when $\tJH /W\approx 0.33$.
For $\tJH /W < 0.33$, $\TR (p)$ vanishes at a single filling that corresponds to $q=0.5$,
meaning that there is no SRO.  For $0.5 > \tJH /W > 0.33$, $\TR (p)$ is nonzero
for all fillings.  As can be seen in Fig.8(a) for $\tJH /W=0.375$, the correlation parameter
for $0.5 > \tJH /W > 0.33$ jumps from $q=0$ in the FM phase to a nonzero value
$0.5 < q < 1$ in the SRO phase.  For $\tJH /W > 0.5$, one of the ordered phases always has a
higher transition temperatures than the SRO states and $q$ jumps from 0 to 1 at the FM/AF boundary.
The deviation between the numeric and weak-coupling results in Fig.8(c) as $p\rightarrow 1$
is quite expected:  in the limit $p\rightarrow 1$, the weak-coupling result for the
SRO transition temperature diverges, thereby invalidating the condition $T\ll \tJH $ under which
it was derived.

The magnetic instabilities of the DE model including SRO states are mapped in Fig.7(b).  Notice
that the region formerly labeled NM in Fig.7(a) is now occupied by SRO states.
The SRO phase also occupies the formerly AF region at very small $\tJH /W$.
As anticipated above, the NM curve (where $\TR (p)=0$ and $q=1/2$) intersects the FM phase boundary
at $\tJH /W \approx 0.33$.  This curve separates a SRO region with AF correlations
($1 > q > 1/2$) on the left from a SRO region with FM correlations ($1/2 > q > 0$) on the right.

\section{Magnetic Phase Diagram of the Double-Exchange Model}

The presence of a continuous range of SRO states with correlation parameter $q$
obviously complicates the magnetic phase diagram of the
DE model and the issue of PS.  Because the weak-coupling behavior of the
DE model has been the subject of some debate \cite{yun:98,nag:00,cha:00,aus:01},
we shall treat this limit analytically.  We then provide numerical results for the
interacting DOS that explain the stability of the SRO phase.  Finally, we
present the $T=0$ phase diagram of the DE model, both with and without the SRO states.

\subsection{Weak-coupling limit}

In their numerical work at $T=0$, Chattopadhyay {\em et al.} \cite{cha:00} obtained a
very complex phase diagram, with a range of SRO states above $\tJH /W \approx 1/16$ and
PS between AF and FM phases for $0.1 < p < 1$ below this value.
For small $\tJH /W$, a pure FM phase was found to be stable for $0 < p \le 0.1$.
We now show that PS diappears in the exactly solvable limit of zero temperature and weak coupling.

At zero temperature, the FM, AF, and SRO energies are evaluated after
once again parameterizing $\uGr_0^{(\al )}(z=i\nu +\mu )^{-1}$
by Eq.(\ref{g0}).  Then, $R(z)$ is obtained from the quartic equation
\begin{equation}
\label{Rn4}
R(z+R)\Bigl\{ (z+2(1-q)R)^2-\tJH^2 \Bigr\} +\ds\frac{W^2}{16}(z+2(1-q)R)^2 = 0,
\end{equation}
which differs from Eq.(\ref{Rl}) (except when $q=1/2$) because we are now working
in the broken-symmetry phase.  Similarly, $Q^{(\uparrow )}(z)=-Q^{(\downarrow )}(z)$ is given by
\begin{equation}
Q^{(\uparrow )}(z)=(2q-1)\frac{M\tJH R(z)}{z+2(1-q)R(z)},
\end{equation}
which is likewise valid for any $\tJH $.

The potential energy can generally be written as
$\langle V\rangle /N =-2\tJH M \langle s_{0z}\rangle $,
where the expectation value of the electronic spin on site $0$ for any temperature is
\begin{equation}
\langle s_{0z}\rangle =\ds\frac{1}{4\pi }\iv \re G^{(\uparrow )}(z)_{\alpha \beta }\sigma^z_{\beta \alpha }
=\ds\frac{8}{\pi W^2(2q-1)}\iv Q^{(\uparrow )}(z).
\end{equation}
For any $\tJH $, the potential energy may be written as
\begin{equation}
\label{pots}
\ds\frac{1}{N} \langle V(p,q) \rangle = -\frac{16\tJH^2 }{\pi W^2}\iv \frac{R(z)}{z +2(1-q)R(z)}.
\end{equation}
Of course, the NM potential energy is obtained by setting $q=1/2$.

Second-order results for $\langle V (p,q) \rangle /N$ are provided in Appendix A.  For small $\tJH /W$,
$\langle s_{0z} \rangle $ is of order $\tJH /W$ so the electronic
spins are not frozen at $T=0$.  Bearing in mind that $\TM (p,q=0)=\TC (p)$ and $\TM (p,q=1)=\TN (p)$,
the difference between the potential energies of a FM, AF, or SRO phase and a NM is given by
$\Delta \langle V(p,q)\rangle /N =-3\TM (p,q)$, where both sides are evaluated to order
$\tJH^2 /W$.  So the potential energy is lowered when the transition temperature is positive.

A general expression for the kinetic energy of the SRO phase is proven in Appendix C.  At $T=0$,
Eq.(\ref{king}) becomes
\begin{equation}
\label{kin}
\frac{1}{N}\langle K \rangle=\frac{W^2}{64\pi }\sum_{\al }\iv G^{(\al )}(z)_{\alpha \beta }
\Bigl\{ (1-q)G^{(\al )}(z)_{\beta \alpha }+qG^{(\bal )}(z)_{\beta \alpha }\Bigr\}.
\end{equation}
In contrast to alternative expressions for the kinetic energy of the FM and AF phases \cite{geo:96},
Eq.(\ref{kin}) does not assume any particular wavevector for the ground state.
Independent of $q$, the zeroth-order kinetic energy is given by $E_0(p)/N=-2W/(3\pi ) (1-\delta^2)^{3/2}$,
where $\delta = 2\mu_0 /W$ and $\mu_0$ is the chemical potential evaluated to zeroth-order in $\tJH /W$.
For any $\tJH $, the SRO kinetic energy is given by
\begin{equation}
\label{kins}
\frac{1}{N}\langle K(p,q) \rangle=\frac{16}{\pi W^2}\iv R(z)^2 \Biggl\{ 1+
\ds\frac{(1-2q)\tJH^2}{\bigl(z+2(1-q)R(z)\bigr)^2}\Biggr\}
\end{equation}
and the NM kinetic is recovered by setting $q=1/2$.
It is important to realize that the second-order kinetic energy contains two contributions:
one from the dependence of the integrand on $\tJH /W$ for a fixed $\mu $;  the other from the
dependence of the chemical potential $\mu (p)$ on $\tJH /W$ for a fixed $p$.

Using the condition that $p=1-32/(\pi W^2)\iv \re R(z) $ depends only on the filling and not on
$\tJH $, we may expand $\mu $ in powers of $\tJH /W$ as $\mu =\mu_0 +(\tJH /W)^2 \mu_2+\ldots $.
Then $\mu_0(p)$ is solved from the condition
\begin{equation}
\label{p0}
p=1+\frac{2}{\pi }\biggl\{ \delta \sqrt{1-\delta^2}+\sin^{-1}\delta \biggr\}
\end{equation}
and $\mu_2(p)$ is given in terms of $\mu_0(p)$ by $\mu_2 =W \delta /\bigl\{(1-q)^2+\delta^2 (2q-1)\bigr\}$.
The divergence of $\mu_2$ as $q\rightarrow 1$ and $p\rightarrow 1$ (or $\delta \rightarrow 0$)
signals the opening of a gap in the interacting, AF DOS as discussed further below.

Incorporating the change in chemical potential, the result for the second-order, $q$-dependent correction
to $\langle K(p,q) \rangle /N$ is given in Appendix A.   To second order in $\tJH /W$, the difference between
the kinetic energies of a FM, AF, or SRO phase and a NM is $\Delta \langle K(p,q)\rangle /N = 3\TM (p,q)/2$.
Hence, kinetic energy is lost  (the electrons become more confined) when the transition temperature is positive.

Consequently, the net change in energy $E(p,q) = \langle K(p,q) + V(p,q)\rangle $
in the weak-coupling limit is given by $\Delta E(p,q) /N=-3\TM (p,q)/2$,
which includes the FM ($q=0$) and AF ($q=1$) phases as special cases.
The same relation would be obtained for a Heisenberg model $H=-(1/2)\sum_{i,j}J_{ij} \vS_i \cdot \vS_j $
with classical spins within Weiss mean-field theory (which becomes exact in infinite dimensions), so that
$\Delta E/N = -J(\vk =0)S^2/2 = -3\TC /2$.  For large $\tJH/W $, however, the DE value of
$\Delta E(p,q=0)/N$ in the FM phase is proportional to $\tJH $ whereas $\TC $
is proportional to $W$.  So the proportionality between $\Delta E(p,q)/N$ and $\TM (p,q)$ (and the analogy with the
Heisenberg model) only holds in the weak-coupling limit.

In the limit of small $\tJH /W$, the ground state energy is minimized by the same correlation parameter
$q$ that maximizes the transition temperature!  This result is not surprising:  for small $\tJH /W$,
the transition temperature is also small so the correlation parameter that appears at $\TM (p,q)$
also minimizes the energy at $T=0$.  As a consistency check, the results in Appendix A may be used to
show that the derivative of the energy is related to the chemical potential by $dE(p,q)/dp = N\mu$,
where both sides are expanded to second order in $\tJH /W$.

\subsection{Interacting Density-of-States}

To better interpret the weak-coupling results, we construct the interacting DOS per spin
\begin{equation}
N(\omega ,q)=-\frac{1}{4\pi }\sum_{\al } \im G^{(\al )}(z \rightarrow \omega +i\delta )_{\alpha \alpha }
=\frac{16}{\pi W^2}\im \Bigl\{  R \Bigr\}_{z \rightarrow \omega +i\delta }.
\end{equation}
As $\tJH /W \rightarrow 0$, $N(\omega ,q)$ reduces to the bare DOS $N_0(\omega )$ for any $q$.
The AF DOS can be solved analytically for all $\tJH $:
\begin{equation}
N(\omega ,q=1) = \frac{8\vert \omega \vert }{\pi W^2}\re \sqrt{\frac{ W^2/4+\tJH^2 -\omega^2 }
{ \omega^2 -\tJH^2 }}
\end{equation}
which vanishes for $\vert \omega \vert  < \tJH $ and $\vert \omega \vert > \sqrt{W^2/4+\tJH^2}$.
Hence, $N(\omega ,q=1)$ contains a square-root singularity on either side of a gap with magnitude $2\tJH $.
As $\tJH /W \rightarrow \infty $, the width of each side-band narrows like $W^2/8\tJH $.
It is also possible to explicitly evaluate the DOS at $\omega =0$ for any $q$:
$N(\omega =0,q)=\bigl(8/\pi W^2 (1-q)\bigr)\re \sqrt{\tJH_c^2 -\tJH^2 }$,
which vanishes for $\tJH > \tJH_c \equiv W(1-q)/2$.  For a NM, the critical value of $\tJH $
required to split the DOS is $\tJH_c = W/4$, as found earlier \cite{che:03}.  For an AF at $T=0$,
$\tJH_c=0$ since a gap forms for any nonzero coupling constant.  For a SRO phase, $\tJH_c> 0$.

\begin{figure}
\includegraphics *[scale=0.6]{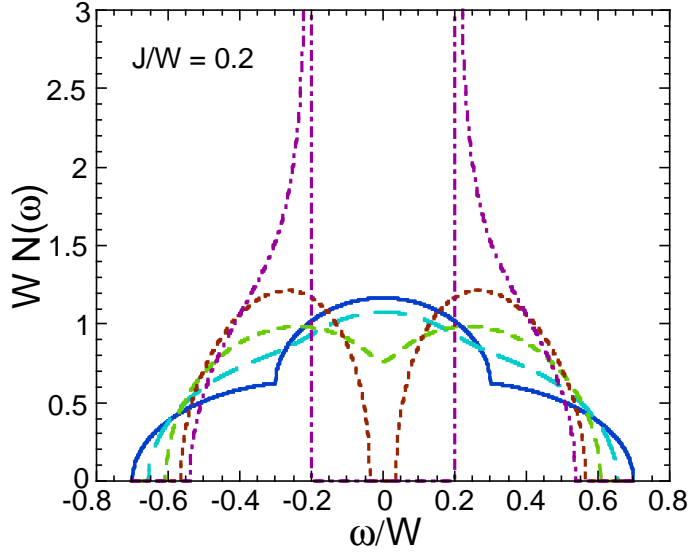}
\caption{
The interacting DOS $N(\omega ,q)$ (normalized by $1/W$) versus $\omega /W$
for $\tJH /W=0.2$ in the FM or $q=0$ (solid, blue), SRO phase with $q=1/4$ (long dash, light blue), NM or $q=1/2$ (dash, green),
SRO phase with $q=3/4$ (small dash, red) and AF or $q=1$ (dot dash, purple) phases.
}
\end{figure}

Taking $\tJH /W=0.2$, we numerically solve for $R(z)$ and plot the interacting DOS
versus $\omega $ for $q=0, 1/4, 1/2, 3/4,$ and 1 in Fig.9.  The FM DOS has kinks at
$\omega =\pm (W/2-\tJH )$.  For energies between the kinks, both up- and down-spin states appear;
on either side of the kinks, the bands are fully spin-polarized.  These kinks disappear in a SRO state
with $q>0 $ due to the absence of long-range magnetic order.  The width of the DOS
shrinks as $q$ increases from 0 to 1.  These results agree with the second-order change to the
chemical potential given above.  In fact, the divergence of the AF result for $\mu_2$ as
$p\rightarrow 1$ is caused by the appearance of a gap in the AF DOS for any nonzero $\tJH $.

For any temperature, coupling constant, and correlation parameter, the total energy
of the DE model may be written as an integral over the interacting DOS:
\begin{equation}
\label{EN}
\frac{1}{N}E(p,q) =2\int d\omega \, \omega f(\omega )N(\omega ,q),
\end{equation}
where $f(\omega )=1/(\exp (\beta (\omega -\mu))+1)$ is the Fermi function and
$\mu =\mu (p)$.  At $T=0$, this relation may be expressed in the
compact form
\begin{equation}
\frac{1}{N}E(p,q)=-\frac{1}{\pi }\iv \Biggl\{ 1 +\frac{16zR(z)}{W^2} \Biggr\},
\end{equation}
which agrees with Eqs.(\ref{pots}) and (\ref{kins}).

From these results, it is not hard to understand why a SRO phase can have a lower energy
than an AF.  Due to the narrowing of the AF DOS, the AF energy may be higher than
that of a SRO state with $\tJH < \tJH_c(q) $ and no energy gap.  If the chemical potential
of the AF phase lies sufficiently far from the energy gap, then the broader DOS of a SRO phase
with $\tJH < \tJH_c$ will favor that state over an AF.  As $p\rightarrow 1$, this
condition is impossible to satisfy for arbitrarily small $\tJH$ and the AF becomes stable.

For a given $p<1$, the expansions of the kinetic and potential energies in powers of
$(\tJH /W)^2$ are valid when $\tJH < \tJH_c(q)$ so that the DOS of a SRO state with correlation
parameter $q$ is not gapped.  Conversely, for a fixed $\tJH $, the expansions
fail sufficiently close to $p=1$ and $q=1$ that a gap develops in $N(\omega ,q)$.

\subsection{Ground-state phase diagram and phase separation}

When only FM, AF, and NM states are considered, PS occurs due to the discontinuity in the
chemical potential at the FM/NM and NM/AF phase boundaries.  Since these discontinuities
become weaker as $\tJH /W$ decreases, PS disappears as $\tJH /W\rightarrow 0$.

\begin{figure}
\includegraphics *[scale=0.6]{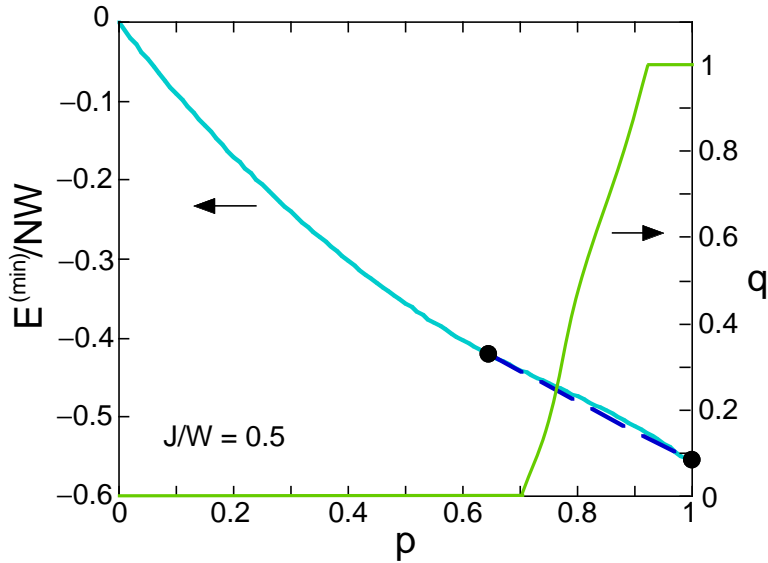}
\caption{
The $T=0$ energy $\Etmin (p)$ and the value of $q$ that minimizes the energy versus filling
for $\tJH/W=0.5$.  A Maxwell construction (given by the dots and dashed line) between fillings
$p_1=0.647$ and $p_2=1$ demonstrates the existence of PS.
}
\end{figure}

Once the SRO states are introduced, the physical energy $\Etmin (p)$ is obtained by minimizing
$E(p,q)$ with respect to $q$ for a fixed $p$.
Although the chemical potential is a continuous function of filling, PS still occurs for non-zero
$\tJH $ due to the formation of an energy gap in the AF state, so that $(1/N)d\Etmin /dp\vert_{p=1^-} =-\tJH $.
A necessary condition for PS between fillings $p_1 < 1$ and $p_2 =1$ is that the second derivative
of the energy $\Etmin (p)$ must change sign for some filling $p^{\star }$ between
$p_1$ and $p_2$.  But if $\Etmin (p)$ is expanded in powers of $(\tJH /W)^2$, then the condition
that $d^2\Etmin (p)/dp^2\vert_{p^{\star }}=0$ may be written
\begin{equation}
\label{Esec}
\frac{\pi }{8\sqrt{1-\delta^2}}\sim \ds\frac{\tJH^2}{W^2 \delta^2 }\Biggl\vert_{p^{\star }},
\end{equation}
where $\delta = 2\mu_0 /W$ and $\mu_0$ is the chemical potential evaluated to zeroth-order
in $\tJH /W$.
Since the value $q^{\star }$ of $q$ at $p^{\star }$ is less than 1, the expansion in powers of $(\tJH /W)^2$ is
valid for $\tJH < \tJH_c(q^{\star })$, as discussed previously.  When $\tJH /W\rightarrow 0$,
the right-hand side of Eq.(\ref{Esec}) vanishes so the condition for PS cannot be met in the limit
of vanishing $\tJH /W$ except as $\delta \rightarrow 0$ or $p^{\star }\rightarrow 1$.  Near $p=1$,
$\delta \approx (1-p)\pi/4$
so Eq.(\ref{Esec}) also implies that the difference $p_2-p_1$ grows linearly with $\tJH/W$.
These results contradict Chattopadhyay {\em et al.} \cite{cha:00}, who seem to use a
Maxwell construction involving the difference $\Etmin (p)-E_0(p)$ rather than the total
minimum energy $\Etmin (p)$.

To show this more explicitly, we have used $\Etmin (p)$ to find the PS region
through Maxwell constructions between fillings $p_1 < 1$ and $p_2=1$.
One such Maxwell construction is pictured in Fig.10, where a phase mixture along the
dashed line provides a lower energy than the pure phase on the solid curve.
For $\tJH /W=0.5$, the SRO states are bypassed and PS occurs between a FM with $p_1=0.647$
and an AF with $p_2=1$.

\begin{figure}
\includegraphics *[scale=0.6]{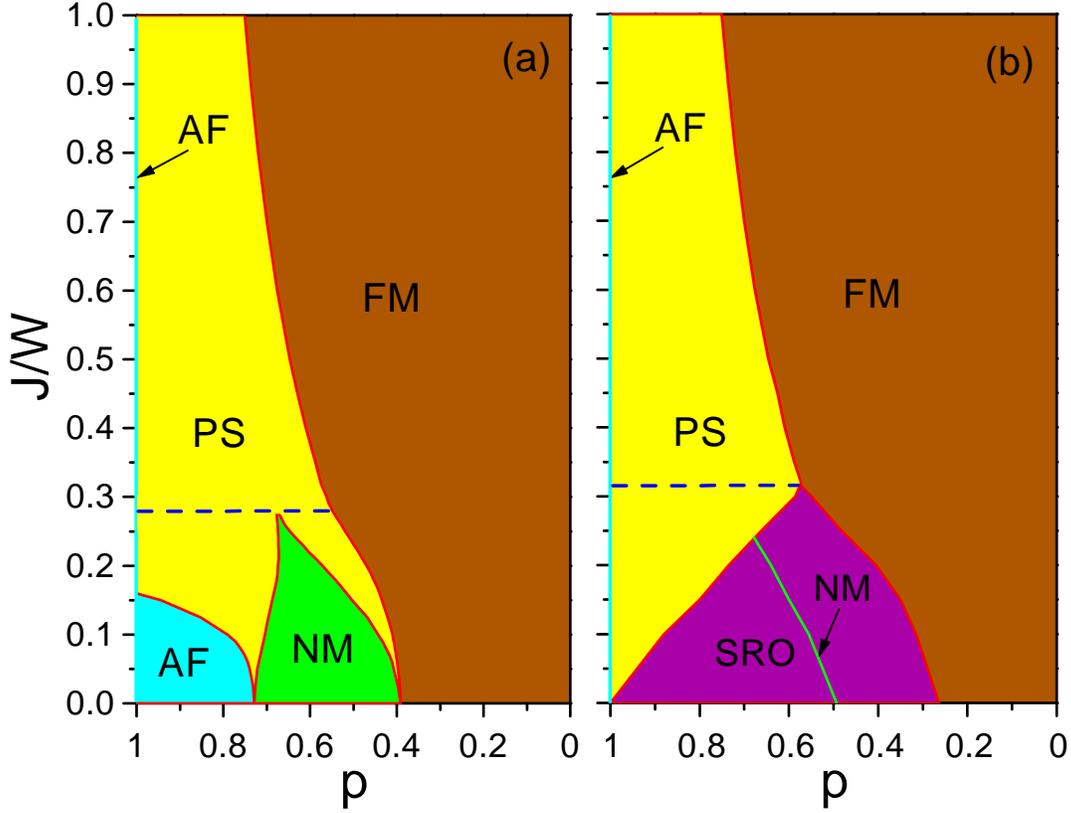}
\caption{
The $T=0$ magnetic phase diagram of the DE model (a) assuming only FM, AF, and NM phases and (b)
also including the SRO phase.
}
\end{figure}

The $T=0$ phase diagram of Fig.11(a) assumes that only the NM, FM, and AF phases are allowed.
Without the SRO state, the ground-state phase diagram is fairly complicated.  As implied by the
argument above, PS disappears in the limit of vanishing $\tJH /W$ but the PS regions grow with
increasing $\tJH /W$.  Two PS regions appear around the fillings of the AF/NM and NM/FM
phase boundaries for $\tJH/W \ll 1$.  When $0 < \tJH /W \le 0.16$, the lower PS region in Fig.11(a)
mixes an AF with $p<1$ and a NM.  For $0.16 < \tJH /W < 0.28$, an AF with $p < 1$ is
unstable and the lower PS region mixes an AF with $p=1$ and a NM.  For $\tJH /W > 0.28$,
PS occurs between a FM and an AF with $p=1$.

Surprisingly, the $T=0$ phase diagram of Fig.11(b) becomes much simpler once the SRO states are included.
For $\tJH /W < 0.33$, a SRO state is stable over a range of fillings that diminishes
with increasing $\tJH $.  Near half-filling, PS is found between a SRO state with $0 < q < 1$ and
an AF with $p=1$.  In agreement with the above discussion, the PS region grows
linearly with $\tJH /W$.  For $\tJH /W> 0.33$, the SRO phase is skipped and PS occurs
between a FM with $q=0$ and an AF with $p=1$.  The special value $\tJH /W \approx 0.33$
agrees with the coupling where the NM line intersects the FM boundary in the instability map of Fig.7(b).
It also corresponds to the coupling where the AF instability covers the widest range of fillings.

The PS region in Figs.11(a) and (b) diminishes with increasing $\tJH /W$.  In the strong-coupling regime,
our results are in qualitative agreement with other authors \cite{fur:95,yun:98}.  But for $\tJH /W=2$,
we obtain PS over a somewhat larger range of fillings than obtained in Ref.\cite{fur:99}.  Aside from the
behavior at very small $\tJH /W$, the phase diagram of Fig.11(b) agrees rather well with
that presented by Chattopadhyay {\em et al.} \cite{cha:00}.  In particular, those authors also
found that the SRO phase disappears when $\tJH /W$ exceeds about 0.3.

Together, Figs.7 and 11 provide a bird's eye view of the magnetic instabilities and phase evolution
in the DE model.  In Figs.7(a) and 11(a), a triangle of phase space remains NM for all temperatures
down to $T=0$;  in Figs.7(b) and 11(b), a somewhat larger triangle of phase space remains in the SRO
phase down to $T=0$.  Keep in mind, however, that the ordered phases of Fig.7 may be bypassed when
$\{ \tJH /W,p\}$ lies in the PS region of Fig.11.  For example, Furukawa \cite{fur:99}
found that for $\tJH /W=2$, an AF with $p < 1$ was excluded by PS between an AF with
$p=1$ and a NM.

\section{Conclusions}

We have used DMFT to evaluate the magnetic phase instabilities and ground-state
phase diagram of the DE model.  Surprisingly, a SRO phase may have a higher transition
temperature than the long-range ordered AF and FM phases.  The ground-state phase
diagram of the DE model actually simplifies in the presence of the SRO states.

At first sight, the stability of disordered NM and SRO ground states may be hard to understand.  After all,
the presence of a disordered state at $T=0$ contradicts Nernst's theorem, which requires
that the entropy of the local moments be quenched at zero temperature.  However, entropic ground states
appear quite frequently in infinite-dimensional calculations.  Within the DMFT of the Hubbard
model, AF ordering on a hypercubic lattice at half filling can be frustrated by hopping between
sites on the same sub-lattice \cite{geo:96}.  This additional hopping  produces a NM Mott
insulator at $T=0$ for large $U/W$, again in violation of Nernst's theorem.
Also on the hypercubic lattice, the DE model has a NM ground state extending all the way to
$p=0$ when $\tJH /W$ is sufficiently small \cite{nag:00}.  Although they did not study
such states in detail, Nagai {\em et al.} \cite{nag:00} detected the signature of incommensurate states
for weak coupling.  Therefore, many-body theories in infinite
dimensions are frequently unable to lower the symmetries of disordered
ground states.  Of course, DMFT cannot access phases described by a non-local
order parameter. It should also be noted that Nernst's theorem only applies rigorously to
quantum systems and not to a DE model with classical local spins.

For small $\tJH /W$, the energy of the entropic SRO state is lowered below the energies
of the FM and AF phases by the RKKY interaction \cite{van:62}, which generates competing
FM and AF interactions \cite{aus:01}.  As the dimension is lowered, the SRO state
may evolve into the state with incommensurate correlations (IC) obtained in one- and
two-dimensional Monte-Carlo simulations by Yunoki {\em et al.} \cite{yun:98}.
Replacing ``SRO'' by ``IC'', the phase diagram of Fig.11(b) bears a striking resemblance to the
phase diagrams of Ref.\cite{yun:98}.  In agreement with our results, Yunoki {\em et al.} found that
the short-range correlations in the IC phase change from AF to FM as $p$ decreases.
The phase diagram of Fig.11(b) also looks quite similar to the three-dimensional phase
diagram obtained by Yin \cite{yin:03}, who compared the energies of the AF and FM phases
with the energy of a two-dimensional spiral phase carrying momenta along the $(111)$ direction.
Hence, the infinite-dimensional phase diagram of Fig.11(b) is closely related to the
results of finite-dimensional calculations.

On a translationally-invariant hypercubic lattice \cite{geo:96} in infinite dimensions, the SRO
phase may map onto an incommensurate spin-density wave (SDW) with a definite wavevector.
In that case, the transition temperature of the SDW could be obtained from either the coupled Green's
functions or the magnetic susceptibility.  However, there are no indications for SDW ordering
in the limited studies of the DE model on an infinite-dimensional hypercubic lattice \cite{nag:00} and
we consider such a possibility to be remote.  Much more likely, the SRO phase would map onto a disordered
state with short-range magnetic order and broad maxima at wavevectors $\vq $ associated
with the nesting of the Fermi surface, as found in Monte-Carlo simulations of one- and two-dimensional
hypercubic lattices \cite{yun:98}.

Bear in mind that DMFT studies of the DE model on a hypercubic lattice sacrifice a key advantage of the
Bethe lattice:  the possibility of obtaining analytic results due to the relationship, Eq.(\ref{sc}), between
the bare and full Green's function.  For example, both sides of the expression $\Delta E(p,q) /N=-3\TM (p,q)/2$
were evaluated independently to order $J^2/W$.  This gives us great confidence
in our results for the energy and critical temperature in the weak-coupling limit.
Such analytic results are not possible on a hypercubic lattice.  Indeed, pathological results
have been found \cite{che:03} on a hypercubic lattice due to the asymptotic freedom of
the quasiparticles in the tails of the Gaussian DOS.  In most instances, more physical results are
obtained on a Bethe lattice, where the bounded DOS more closely resembles the DOS in
two and three dimensions.

Working on a Bethe lattice has allowed us to clarify the behavior of the DE model at weak coupling.
We have demonstrated both analytically and numerically that PS disappears as $\tJH /W\rightarrow 0$.
Whereas the FM phase is stable for $p$ below about 0.26, a SRO state is stable above.  As $p\rightarrow 1$,
the SRO phase evolves into the long-range ordered AF phase.  The transition temperature of
the SRO phase exceeds that of the FM and AF phases for $0.26 < p < 1$.

Whether the SRO phase is a new kind of spin glass or a spin liquid can only be resolved
by future studies of the static and dynamic susceptibilities.  An intriguing possibility is that the local moments are
frozen within a spin-glass state but that the electronic spins are subject to quantum fluctuations
within a spin-liquid state.  Two of the most important unresolved questions about spin glasses are
whether there exists a true thermodynamic spin-glass transition \cite{bit:97}
and whether a model without quenched disorder can support a spin glass \cite{sch:00}.
We have answered those questions for the DE model on an infinite-dimensional Bethe lattice.
Above $\TM (p,q)$, the NM state is characterized by the absence of SRO around any site.  Below $\TM (p,q)$,
short-range magnetic order develops with a correlation length
$\xi (q) = -a/\log \vert 2q-1\vert $ and the entropy, while still extensive, is reduced.  Although not marked by a
divergent susceptibility, the transition at $\TM (p,q)$ is characterized by the development of short-range
magnetic order and a reduction in the entropy.  Future work may establish connections between the SRO
phase and the spin-glass solution to the random Ising model on a Bethe lattice \cite{spgbl}, where the
Edwards-Anderson order parameter can be explicitly constructed.

The strength of DMFT applied to DE-like models with a Kondo coupling
$2\tJH \hm_i\cdot \vs_i$ between the local-moment spin $\vS_i =S\hm_i $
and the electron spin $\vs_i$ is that the Fermion degrees of freedom
in the local action may be integrated exactly, leaving a function only
of the local moment direction $\hm_i $.  Consequently, formally exact
results are possible for a range of models where the local moments
are treated classically.  This stands in stark contrast to the case for
the Hubbard model \cite{geo:96}, where the action cannot be analytically integrated
and quantum Monte Carlo must be used to solve the local problem.
We should also mention that the classical treatment of the local spins is
more questionable in the manganites, where $S=3/2$, than in Mn-doped
GaAs, where $S=5/2$.

The discovery of a SRO solution to the DE model in infinite dimensions has wide
ramifications.   Our results open the door for future studies of the disordered ground-state
of the DE model unencumbered by the numerical complexities of Monte-Carlo
simulations \cite{yun:98} in finite dimensions.

It is a pleasure to acknowledge helpful conversations with Profs. Elbio Dagotto, Jim Freericks,
Douglas Scalapino, and Seiji Yunoki.  This research was sponsored by the U.S. Department of Energy
under contract DE-AC05-00OR22725 with Oak Ridge National Laboratory, managed by UT-Battelle, LLC and
by the National Science Foundation under Grant Nos. DMR-0073308, DMR-0312680,
EPS-0132289 and EPS-0447679 (ND EPSCoR).  A portion
of this research was conducted at the Center for Nanophase Materials Sciences, which is sponsored at
Oak Ridge National Laboratory by the Division of Scientific User Facilities, U.S. Department of Energy.

\appendix

\section{Weak-coupling results}

This appendix presents the weak-coupling results for the SRO transition temperature
and energy.  For low temperatures, $\sum_n g(\nu_n)\rightarrow (1/2\pi T)\int d\nu \, g(\nu )$ so that
\begin{equation}
\label{Tsga}
\TM (p,q) =-\frac{\tJH ^2(2q-1)}{3\pi }\iv \frac{R(v)}{\bigl(v+R(v)\bigr)^2\bigl(v+2(1-q)R(v)\bigr)
-\tJH^2(v+2(2-q)R(v)/3)},
\end{equation}
where $v=i\nu +\mu $.  In the weak-coupling limit $T \ll \tJH \ll W$, this becomes
\begin{equation}
\TM (p,q) =-\frac{\tJH ^2(2q-1)}{3\pi }\iv \frac{R(z)}{\bigl(z+R(z)\bigr)^2\bigl(z+2(1-q)R(z)\bigr)},
\end{equation}
where $z=i\nu +\mu_0$ and $R(z)$ falls off
like $W^2/z$ for large $\vert z\vert $.  The zeroth-order chemical potential $\mu_0$ is defined in terms
of the filling $p$ by Eq.(\ref{p0}).  Using complex analysis and carefully treating the branch cut between
$\pm W/2$ along the real axis, we find that
\begin{equation}
\ds\frac{\TM (p,q)W}{\tJH^2} =-\ds\frac{32}{3\pi }\int_{\delta }^1 dy\, y\sqrt{1-y^2}
\, \ds\frac{y^2-(3-2q)/4}{y^2+b},
\end{equation}
with $\delta = 2\mu_0 /W$ and $b=(1-q)^2/(2q-1)$.
Clearly, $\TM (p,q)$ vanishes as $q\rightarrow 1/2$ or $b\rightarrow \infty $.

In the regime $0 \le  q < 1/2$, the integral for $\TM (p,q)$ with $b < 0$ gives
\begin{eqnarray}
&\ds\frac{\TM (p,q)W}{\tJH^2}=\ds\frac{8}{9\pi (1-2q)}\Biggl\{
3\sqrt{1-\delta^2}-4(1-\delta^2)^{3/2}(1-2q) \nonumber \\ &
+\ds\frac{3q}{2\sqrt{1-2q}} \Biggl[
\Tan^{-1} \Biggl( \ds\frac{(1-q)\vert \delta \vert -\sqrt{1-2q}}{\sqrt{1-\delta^2}q}\Biggr)
-\Tan^{-1} \Biggl( \ds\frac{(1-q)\vert \delta \vert +\sqrt{1-2q}}{\sqrt{1-\delta^2}q}\Biggr)
\Biggr]\Biggr\}.
\end{eqnarray}
This result can be extended to the regime $1/2 < q \le 1$ by using the identity
\begin{eqnarray}
&\ds\frac{1}{\sqrt{1-2q}} \Biggl[
\Tan^{-1} \Biggl( \ds\frac{(1-q)\vert \delta \vert -\sqrt{1-2q}}{\sqrt{1-\delta^2}q}\Biggr)
-\Tan^{-1} \Biggl( \ds\frac{(1-q)\vert \delta \vert +\sqrt{1-2q}}{\sqrt{1-\delta^2}q}\Biggr)
\Biggr] \\ &
=-\ds\frac{2}{\sqrt{2q-1}}\log \Bigg[ \ds\frac{q+\sqrt{1-\delta^2}\sqrt{2q-1}}
{\sqrt{\delta^2(2q-1)+(1-q)^2}}\Biggr]
\end{eqnarray}
These expressions yield Eq.(\ref{TCwc}) for $\TC $ and Eq.(\ref{TNwc}) for $\TN $
as $q\rightarrow 0$ and $q\rightarrow 1$, respectively.

Using Eq.(\ref{kin}), it is straightforward to obtain the kinetic energy in the weak-coupling
limit:
\begin{eqnarray}
&\ds\frac{1}{N}\langle K(p,q) \rangle = \ds\frac{1}{N}E_0(p)
+\ds\frac{2\tJH^2}{\pi W (1-2q)}\Biggl\{ 2\sqrt{1-\delta^2 } +\ds\frac{q}{\sqrt{1-2q}}
\nonumber \\ &
\times \Biggl[
\Tan^{-1} \Biggl( \ds\frac{(1-q)\vert \delta \vert -\sqrt{1-2q}}{\sqrt{1-\delta^2}q}\Biggr)
-\Tan^{-1} \Biggl( \ds\frac{(1-q)\vert \delta \vert +\sqrt{1-2q}}{\sqrt{1-\delta^2}q}\Biggr)
\Biggr]\Biggr\}.
\end{eqnarray}
Of course, the NM kinetic energy is obtained by taking the limit $q\rightarrow 1/2$,
with the result $\langle K (p,q=1/2)\rangle /N = E_0(p)/N +(16\tJH^2/3\pi W)(1-\delta^2)^{3/2}$.
The difference between the $q$-dependent and NM kinetic energies
in the weak-coupling limit is then given by $\Delta \langle K(p,q) \rangle /N = 3\TM (p,q)/2$.

Likewise, it is straightforward to evaluate the potential energy to order $(\tJH /W)^2$,
with the result:
\begin{eqnarray}
&\ds\frac{1}{N}\langle V(p,q) \rangle =-\ds\frac{4\tJH^2}{\pi W(1-2q)}\Biggl\{
2\sqrt{1-\delta^2 } +\ds\frac{q}{\sqrt{1-2q}}
\nonumber \\ &
\times \Biggl[
\Tan^{-1} \Biggl( \ds\frac{(1-q)\vert \delta \vert -\sqrt{1-2q}}{\sqrt{1-\delta^2}q}\Biggr)
-\Tan^{-1} \Biggl( \ds\frac{(1-q)\vert \delta \vert +\sqrt{1-2q}}{\sqrt{1-\delta^2}q}\Biggr)
\Biggr]\Biggr\}.
\end{eqnarray}
In the limit $q\rightarrow 1/2$, the NM result is $\langle V(p,q=1/2) \rangle /N = -(32\tJH^2/3\pi W)(1-\delta^2)^{3/2}$.
Hence, the difference between the $q$-dependent and NM potential energies in the weak-coupling limit is given
by $\Delta \langle V(p,q) \rangle /N = -3\TM (p,q)$.  It follows that the change in total energy is given by
$\Delta E (p,q)/N =-3\TM (p,q)/2$, which includes the FM ($q=0$) and AF ($q=1$) phases as special cases.

\section{Self-consistency relation for the SRO states}

Following the cavity method of Ref.\cite{geo:96}, the self-consistency relation on the Bethe lattice
between the bare and full Green's functions may be written
\begin{equation}
\label{gt}
\uGr_0(i\nu_n)^{-1}=z_n\uI  -t^2 \sum_i' \uGr (i\nu_n)_{ii},
\end{equation}
where $t^2=W^2/16\zc $ and $\uGr (i\nu_n )_{ii}$ is the local Green's function on site $i$.
The sum on $i$ is restricted to nearest neighbors of site 0.
If the spin on site $0$ is ordered along the $z$ axis, then we can write the local Green's function
on site $0$ as
\begin{equation}
\uGr (i\nu_n )_{00 } =
\left(
\begin{array}{cc}
g_{\uparrow n} & 0 \\
0 & g_{\downarrow n}\\
\end{array} \right) .
\end{equation}
The creation and destruction operators on site $i$ are taken to refer to a quantization axis that is rotated
by angle $\theta_i $ (rotation axis in the $xy$ plane at an angle $\phi_i$ from the $x$ axis) with respect to $z$.  Then
$\uGr (i\nu_n)_{ii} = \underline{R}(\theta_i, \phi_i) \, \uGr (i\nu_n)_{00}\,
\underline{R}(\theta_i ,\phi_i )^{\dagger }$ is given by
\begin{equation}
\uGr (i\nu_n )_{ii } =
\left(
\begin{array}{cc}
\cos^2 (\theta_i/2) g_{\uparrow n} + \sin^2 (\theta_i/2) g_{\downarrow n} &
i\sin \theta_i  e^{-i\phi_i } \bigl(g_{\uparrow n}-g_{\downarrow n}\bigr)/2 \\
-i\sin \theta_i  e^{i\phi_i } \bigl(g_{\uparrow n}-g_{\downarrow n}\bigr)/2&
\cos^2 (\theta_i/2) g_{\downarrow n} + \sin^2 (\theta_i/2) g_{\uparrow n} \\
\end{array} \right) .
\end{equation}
Summing over all $\zc $ neighbors of site 0 and assuming that $\phi_i$ is random so that
$(1/\zc )\sum_i' \exp (\pm i\phi_i)=0$, we find
\begin{equation}
\frac{1}{\zc } \sum_i' \uGr (i\nu_n )_{ii } = (1-q)
\left(
\begin{array}{cc}
g_{\uparrow n} & 0 \\
0 & g_{\downarrow n}\\
\end{array} \right) +q
\left(
\begin{array}{cc}
g_{\downarrow n} & 0 \\
0 & g_{\uparrow n}\\
\end{array} \right),
\end{equation}
where $q=(1/\zc ) \sum_i' \sin^2 (\theta_i/2)$.   Upon noting that the two Green's function matrices
above are spin-reversed, we obtain Eq.(\ref{scsg}).

\section{Kinetic energy}

Previous forms for the kinetic energy utilized the wavevector $\vq =0$ or $\vQ$ of the FM or AF
phase \cite{geo:96}.  In this appendix, we develop an expression for the kinetic energy that also applies to the
SRO states, where the wavevector is not defined.  Begin with the relation
\begin{equation}
\ds\frac{1}{N}\langle K \rangle = -\ds\frac{t}{2Z_f} \sum'_{i}\int \Pi_i\,  d\Omega_{\hm_i} \Tr
\Bigl( (c_{0\alpha }^{\dagger }c_{i\alpha } +c_{i\alpha }^{\dagger }c_{0\alpha })
e^{-S}\Bigr),
\end{equation}
where $Z_f =\int \Pi_i \, d\Omega_{\hm_i }\Tr \Bigl( \exp (-S) \Bigr) $ and the sum over $i$ is restricted
to nearest neighbors of site 0.  The action $S=\Delta S +S_0 +S^{(0)}$ is separated into terms that involve
site $0$ and the action $S^{(0)}$ of the cavity \cite{geo:96} with site 0 removed.  The kinetic part of the
action containing site 0 is
\begin{equation}
\Delta S = -t\int_0^{\beta }d\tau \sum'_{j}\Bigl( c_{0\alpha }^{\dagger }(\tau )c_{j\alpha }(\tau )
+c_{j\alpha }^{\dagger }(\tau )c_{0\alpha }(\tau )\Bigr).
\end{equation}
Performing perturbation theory in the term involving the $(0i)$ bond and using $t=W/4\sqrt{\zc }$, we find
\begin{eqnarray}
\label{king}
&\ds\frac{1}{N}\langle K\rangle =\ds\frac{W^2}{16\zc }\sum'_{i}\int_0^{\beta } d\tau
\langle c_{i\alpha }^{\dagger }(0) c_{i\alpha }(\tau )\rangle
\langle c_{0\alpha }^{\dagger }(\tau ) c_{0\alpha }(0 )\rangle
\nonumber \\
&=\ds\frac{W^2}{32}T\sum_{l,\al } G^{(\al )}(i\nu_l)_{\alpha \beta }\Bigr\{
(1-q)G^{(\al )}(i\nu_l)_{\beta \alpha }+qG^{(\bal )}(i\nu_l)_{\beta \alpha }\Bigl\},
\end{eqnarray}
where the local averages at sites 0 and $i$ are performed using $\seff $ of Eq.(\ref{act}).
Removing the single $(0i)$ bond from $\Delta S$ does not affect those local averages in the
$\zc \rightarrow \infty $ limit.

\end{document}